 \documentclass[leqno]{amsart}

\usepackage{amsmath}
\usepackage{amsthm}
\usepackage{amssymb}
\usepackage{amsfonts}
\usepackage[applemac]{inputenc}
\usepackage{mathrsfs}
\usepackage{pdfsync}

\def\C{{\mathbb C}}
\def\R{{\mathbb R}}
\def\N{{\mathbb N}}
\def\le{\leqslant}
\def\ge{\geqslant}

\newcommand{\im}{\mathrm{Im}}
\newcommand{\e}{\varepsilon}

\DeclareMathOperator{\diver}{div}

\theoremstyle{plain}
\newtheorem{theorem}{Theorem}[section]
\newtheorem{lemma}[theorem]{Lemma}
\newtheorem{corollary}[theorem]{Corollary}
\newtheorem{proposition}[theorem]{Proposition}

\theoremstyle{definition}
\newtheorem{definition}[theorem]{Definition}

\newtheorem{remark}[theorem]{Remark}
\newtheorem*{remark*}{Remark}
\newtheorem*{example}{Example}

\numberwithin{equation}{section}

\begin{document}

\title[Bohmian measures]
{Bohmian measures and their classical limit}
\author[P. Markowich]{Peter Markowich}
\address[P. Markowich]{Department of Applied Mathematics and Theoretical
Physics\\
CMS, Wilberforce Road\\ Cambridge CB3 0WA\\ United Kingdom}
\email{p.markowich@damtp.cam.ac.uk}
\author[T. Paul]{Thierry Paul}
\address[T. Paul]{CNRS and Centre de math\'ematiques Laurent Schwartz, \'Ecole Polytechnique, 91 128 Palaiseau cedex\\ France}
\email{paul@math.polytechnique.fr}
\author[C. Sparber]{Christof Sparber}
\address[C. Sparber]{Department of Applied Mathematics and Theoretical
Physics\\
CMS, Wilberforce Road\\ Cambridge CB3 0WA\\ United Kingdom}
\email{c.sparber@damtp.cam.ac.uk}
\begin{abstract}
We consider a class of phase space measures, which naturally arise in the Bohmian interpretation of quantum mechanics. 
We study the classical limit of these so-called Bohmian measures, in dependence on the scale 
of oscillations and concentrations of the sequence of wave functions under consideration. The obtained results are consequently compared 
to those derived via semi-classical Wigner measures. 
To this end, we shall also give a connection to the theory of Young measures and prove several new results 
on Wigner measures themselves. Our analysis gives new insight on oscillation and concentration effects in the semi-classial regime.
\end{abstract}

\date{\today}


\thanks{This publication is based on work supported by Award No. KUK-I1-007-43, funded by the King Abdullah University of Science and Technology (KAUST). C.S. has been supported 
by the Royal Society via his University research fellowship and P.M. by his Royal Society Wolfson Research Merit Award.}
\maketitle

\tableofcontents

\section{Introduction}

In this work we consider a quantum systems described by a time-dependent wave-function $\psi^\e (t, \cdot) \in L^2(\R^d; \C)$. The dynamics 
of $\psi^\e$ is governed by the linear Schr\"odinger equation
\begin{equation}
\label{schro}
i\e \partial _t \psi^\e  = -\frac{\e^2}{2}\Delta \psi^\e +
V(x)\psi^\e,\quad
\psi^\e (0, x)   = \psi^\e_{0} (x).
\end{equation}
where $x \in \R^d$, $t\in \R $ and $V=V(x)$ a given real-valued potential. Here, we already rescaled all physical parameters, such that only one dimensionless parameter $ \e >0$ remains. We shall mainly be 
interested in the \emph{classical limit} $\e \to 0_+$, and so in the following refer to $\e$ as the \emph{semi-classical parameter}.
In the usual interpretation of quantum mechanics, the wave-function $\psi^\e$ 
yields a probabilistic description of the position of the particle $X(t)\in \R^d$, at time $t \in \R$. More precisely, 
$$
\text {Prob$_{X(t) \in \Omega}$}= \int_ \Omega |\psi^\e(t,x)|^2 dx,
$$
is the probability of finding the particle a time $t \in \R$ within the region $\Omega \subseteq \R^d$. This requires  
the wave function to be normalized $\| \psi^\e(t,\cdot) \|_{L^2} = 1$. 

In more generality, we recall that, although the wave function $\psi^\e$ itself is not a physical 
observable, (real-valued) quadratic quantities of $\psi^\e$ yield probability densities for the respective physical observables. 
Two important examples of such densities, describing the expected values of observables (in a statistical interpretation),
are the \emph{position} and the \emph{current density}, i.e.
\begin{equation}\label{densities}
\rho^\e(t,x)= |\psi^\e(t,x)|^2, \quad J^\e(t,x) = \e \im\big(\overline{\psi^\e}(t,x)\nabla \psi^\e(t,x)\big).
\end{equation}
It is easily seen that if $\psi^\e$ solves \eqref{schro}, then the following conservation law holds
\begin{equation}
\label{con}
\partial_t \rho^\e + \diver J^\e = 0.
\end{equation}
Similarly, one can define the \emph{total energy} of the particle, which is conserved along sufficiently smooth solutions to \eqref{schro}. In our case it is given by
\begin{equation}\label{energy}
E[\psi^\e(t)] =   \frac{\e^2}{2} \int _{\R^d} | \nabla \psi^\e (t,x) |^2 dx + \int_{\R^d} V(x) |  \psi^\e (t,x) |^2 dx, 
\end{equation}
i.e. by the sum of the kinetic and the potential energy.  

The semi-classical regime of quantum mechanics corresponds to situations where $\e \ll 1$. Note that $\e$ corresponds to the typical wave-length of oscillations within the sequence of wave functions $\psi^\e$. In view of 
\eqref{schro}, this is a highly singular asymptotic regime and thus analyzing the limiting behavior of expectation values of physical observables requires analytical care. 
In particular, the limit of the highly oscillatory wave function $\psi^\e$ itself 
is of almost no relevance due to the non-commutativity of weak limits and nonlinear functions. 

The conservation law \eqref{con}, is also a possible starting point of the \emph{Bohmian interpretation} of quantum mechanics
 \cite{BO}
(see also \cite{DuTe} for a broader introduction). To this end 
one introduces the velocity field 
\begin{equation}\label{velocity}
u^\e(t,x):= \frac{J^\e(t,x)}{\rho^\e(t,x)}= \e \im \left(\frac{ \nabla \psi^\e(t, x)}{ \psi^\e(t, x)} \right ) ,
\end{equation}
which is well-defined, expect at nodes, i.e. zeros, of the wave function $\psi^\e$. Ignoring this problem for the moment, the Bohmian dynamics of quantum particles is 
governed by the following system of ordinary differential equations for the macroscopic position vector:
\begin{equation}\label{bohm1}
\dot X^\e(t, x) = u^\e(t,X^\e(t, x)) ,\quad X^\e(0, x) = x\in \R^d.
\end{equation} 
This can be considered as the \emph{Eulerian viewpoint} of 
Bohmian mechanics, with $u^\e$ the associated Eulerian velocity. 
The solution on the Schr\"odinger equation, usually called \emph{pilot-wave}, is hereby used to define a dynamical system 
governing the time-evolution of particles via \eqref{bohm1}. 
In addition one assumes that initially the particle's position is not completely known, but described by the probability distribution $ \rho^\e (0,x)  = |  \psi^\e_{0} (x) |^2 $. This probabilistic 
feature of Bohmian mechanics can be understood as a lack of knowledge about the fine details of the considered quantum mechanical system, 
analogously to the situation in classical statistical mechanics, cf. \cite{DuTe}. 

In order to study the semi-classical limit of Bohmian mechanics, one might hope to analyze directly the limit of the trajectories  $X^\e(t)$ as $\e \to 0_+$. 
To our knowledge, the only rigorous result in this direction has recently been given in \cite{DuRo}, where the authors restrict themselves 
to the case of so-called \emph{semi-classsical wave packets}, see Section 5.3. In the present work we shall consider a more general situation 
which, as $\e \to 0_+$, leads to concentration and oscillation effects (within the particle and current density) not present in the case of \cite{DuRo}. In particular it will become clear from the 
Examples given in Section 5 and the comparison to WKB asymptotics (see Section 6) 
that a direct study of the limiting Bohmian trajectories is very hard and a complete solution of this problem seems out of reach so far 
(cf. \cite{AKST} where the most recent results on the limiting behavior of singularly perturbed 
dynamical systems are proved, albeit in a setting much simpler than ours).
Instead, we shall pass to the corresponding \emph{Lagrangian point of view} on Bohmian dynamics and 
argue that this naturally leads to the introduction of 
a certain class of probability measures on phase space, which we shall call \emph{Bohmian measures}. 
These measures are supported on sub-manifolds (of phase space) 
induced by the graph of the velocity field $u^\e(t, x)$. They consequently evolve via the Bohmian phase space flow and can be shown to be \emph{equivariant} 
with respect to this flow. In addition the first and second moment of these measures yield the correct quantum mechanical position and current density. 
It is therefore natural to consider the classical limit of Bohmian measures in order to analyze the emergence of classical dynamics from 
Bohmian mechanics (see also Remark \ref{rem: measXP} for some preliminary observations regarding the connection to the classical limit of Bohmian trajactories).

The main analytical tool for studying the classical limit of Bohmian measures will be the theory of Young measures, 
which provides information on weak limits of oscillatory sequences of functions.
The obtained limit will then be compared to the well established theory of (semi-classical) 
\emph{Wigner measures} associated to $\psi^\e(t)$, see e.g. \cite{LiPa, GMMP, SMM}. These are phase space measures which, after taking appropriate moments, are  
known to give the \emph{correct} classical limit of (the probability densities corresponding to) physical observables.  
We shall prove that the limiting Bohmian measure of $\psi^\e(t)$ coincides with its naturally associated Wigner measure locally in-time. That is, 
before caustic onset, where the first singularity occurs in the solution of the corresponding 
classical Hamilton-Jacobi equation, see Section 6. Furthermore, we shall argue (by examples) that in general the limiting Bohmian measure differs from the Wigner measure after caustic onset.
In the course of this we shall also prove new results on when Wigner transforms tend to mono-kinetic Wigner measures.

The purpose of the present work is thus twofold: First, to gain some information on the classical limit of Bohmian mechanics (see in particular Section 6). Second, to give further insight 
on oscillation and concentration effects in the semi-classical regime by means of two physically natural, yet again mathematical very 
different, descriptions via phase space measures. Moreover, we believe that our analysis may very well be used as a first building block towards an 
\emph{optimal transportation formulation of quantum mechanics}, by combining our results with those given in \cite{AMGA, GNT} and \cite{GST}, see also Remark \ref{kinremark} below. 

\section{A Lagrangian reformulation of Bohmian mechanics}\label{sec:Lag}

\subsection{Existence of Bohmian trajectories} 
We start with some basic assumptions on the potential $V$. Since in this work we shall not be concerned with  
regularity issues we assume 
\begin{equation}\label{assumptionV}
V \in C^\infty (\R^d; \R) , \quad  V(x) \ge 0. 
\tag{A.1}
\end{equation}
This is (by far) sufficient to ensure that the Hamiltonian operator
\begin{equation}\label{ham}
H^\e= -\frac{\e^2}{2}\Delta + V(x),
\end{equation}
is essentially self-adjoint on $D(H^\e)=C_0^\infty \subset L^2(\R^d;\C)$, cf. \cite{ReSi}. Its unique self-adjoint extension (to be denoted by the same symbol) 
therefore generates a unitary $C_0-$group $U^\e(t) = e^{-it H^\e /\e}$ on $L^2(\R^d)$, which ensures 
the global existence of a unique solution $\psi^\e(t) = U^\e(t) \psi_0$ of the Schr\"odinger equation \eqref{schro}, such that 
$$
\| \rho^\e(t, \cdot) \|_{L^1} \equiv \| \psi^\e(t, \cdot) \|^2_{L^2} = \| \psi^\e_0 \|^2_{L^2},\quad \forall \, t \in \R.
$$
From now on, we shall also impose the following assumption on the initial data:
\begin{equation}\label{assumptionW}
\text{$\psi^\e_0 \in C^\infty (\R^d)$, with $\|\psi^\e_0\|_{L^2}=1$ and $E[\psi_0^\e] < + \infty$, uniformly in $\e$.}
\tag{A.2}
\end{equation}
Since $U^\e(t)$ and $H^\e$ commute, we also have that the total energy is conserved, i.e.
$$
E[\psi^\e(t)]  = E[\psi^\e_0] , \quad \forall \, t \in \R,
$$
and thus,  in view of \eqref{assumptionW}, $E[\psi^\e(t)] $ is uniformly bounded as $\e \to 0_+$ and for all $t\in \R$. 
In addition, the dispersive properties of $U^\e(t)$ together with the assumption \eqref{assumptionV} imply that if $\psi^\e_0 \in C^\alpha (\R^d)$, for $\alpha \ge 0$, then 
$\psi^\e(t, \cdot) \in C^\alpha(\R^d)$ for all times $t \in \R$.  In the following, we denote 
$$
\| f^\e  \|_{H^1_\e}:= \| f^\e  \|_{L^2}  + \| \e \nabla f^\e  \|_{L^2}  
$$
and we say that a sequence $f^\e \equiv \{f^\e\}_{0<\e \le 1}$ is \emph{uniformly bounded} (as $\e\to 0_+$) in $H^1_\e(\R^d;\C)$, if 
$$
\sup_{0<\e \le 1} \| f^\e  \|_{H^1_\e} < + \infty.
$$
Note that the two conservation laws given above, together with \eqref{assumptionV}, \eqref{assumptionW}, imply that  
$\psi^\e(t, \cdot) \in H^1_\e(\R^d)$ uniformly bounded as $\e \to 0_+$ and for all $t\in \R$. Moreover, in view of \eqref{densities} we have
\begin{equation}\label{Jest}
\| J^\e  (t, \cdot)\|_{L^1} \le \e \|  \nabla \psi^\e(t, \cdot) \|_{L^2} \, \| \psi^\e(t,\cdot) \|_{L^2}\le E[\psi_0^\e],
\end{equation}
and we conclude that for all $t\in \R$: $J^\e(t, \cdot)\in L^1(\R^d;\R^d)$ uniformly as $\e\to 0_+$, provided assumptions \eqref{assumptionV} and \eqref{assumptionW} hold.

Next, we recall the main result of \cite{TeTu} (see also \cite{BDGPZ}) on the global existence of Bohmian trajectories.
\begin{proposition}\label{existence}
Let \eqref{assumptionV} , \eqref{assumptionW} be satisfied. Then the map $X^\e_t: x \mapsto X^\e(t,x) \in \R^d$ induced by \eqref{bohm1} exists globally in-time for 
almost all $x \in \R^d$, relative to the measure $\rho_{0} ^\e= |\psi^\e_{0}(x)|^2 dx$ and $X^\e_t \in C^1$ on its maximal open domain. 

Moreover, the probability density $\rho^\e(t, \cdot)$ is the push-forward of the initial density $ \rho_{0} ^\e$ under the map $X^\e_t$, i.e.
\begin{equation*}
\rho^\e(t) = X^\e_t \, \# \,  \rho_{0} ^\e .
\end{equation*}
\end{proposition}
We consequently infer that the Bohmian trajectories, defined through the ordinary differential equation 
\eqref{bohm1}, exist $\rho^\e_{0}-a.e.$ and that for any compactly supported test function
$\sigma \in C_0(\R^d)$ it holds
\begin{equation}\label{push}
\int_{\R^d} \sigma(x) \rho^\e(t,x) dx = \int_{\R^d} \sigma(X^\e(t,x)) \rho^\e_{0} (x) dx.
\end{equation}
This property is also called \emph{equivariance} of the measure $\rho^\e(t, \cdot)$ in \cite{BDGPZ, TeTu}. In addition we may interpret \eqref{push} as a way of giving sense to the solution of the continuity equation
\begin{equation}\label{conservation}
\partial_t \rho^\e + \diver ( \rho^\e u^\e) = 0,
\end{equation}
where $u^\e$ is given by \eqref{velocity}. Due to the possible occurrence of nodes in $\psi^\e(t,x)$, the vector field $u^\e(t,x)$ is in general not Lipschitz in $x$. In fact, not even the 
general existence theory \cite{Am, DiLi} for velocity vector fields which only have a certain Sobolev or BV regularity applies to Bohmian trajectories. 
The property \eqref{push} therefore can only interpreted as a very weak notion of solving the continuity equation \eqref{conservation}.  

\subsection{Bohmian measures on phase space}  
We shall now reformulate Bohmian mechanics, in its (well-known) Lagrangian formulation. To this end we first introduce 
the Lagrangian velocity
$$
P^\e(t, x) = \dot X^\e(t, x),
$$
for which we want to derive an equation of motion. In view of \eqref{bohm1}, we can differentiate $P^\e(t,x)$ $\rho^\e_0 - a.e.$ 
to obtain 
\begin{equation}\label{peq}
\begin{aligned}
\dot P^\e(t, x) = & \ \partial_t u^\e(t, X^\e(t,x)) + \big(\dot X^\e(t,x) \cdot \nabla \big) u^\e(t, X^\e(t,x))\\
= & \ \partial_t u^\e(t, X^\e(t,x)) + \big(u^\e(t,X(t,x)) \cdot \nabla \big) u^\e(t, X^\e(t,x)).
\end{aligned}
\end{equation}
To proceed further, we need an equation for the velocity field $u^\e$.  To this end, we recall the well-known \emph{hydrodynamic reformulation of quantum mechanics}, where one derives 
from \eqref{schro} a closed system of equations for the densities $\rho^\e, J^\e$. Assuming that $\psi^\e$ is sufficiently differentiable, they are found to be (see e.g. \cite{GaMa})
\begin{equation}
\label{qhd}
\left \{
\begin{aligned}
& \, \partial_t \rho^\e + \diver J^\e = 0,\\
& \, \partial_t J^\e + \diver \left(\frac{J^\e \, J^\e}{\rho^\e} \right) + \rho^\e \nabla V = \frac{\e^2}{2} \rho^\e  \nabla \left( \frac{\Delta \sqrt{\rho^\e}}{ \sqrt{\rho^\e}} \right).
\end{aligned}
\right. 
\end{equation}
Under the regularity assumptions on $V$ and $\psi^\e$ stated above, 
the weak formulation of the quantum hydrodynamic equations \eqref{qhd} holds in a rigorous way, i.e. each of the nonlinear terms can be interpreted in the sense of distributions, see \cite[Lemma 2.1]{GaMa}. 
\begin{remark} Let us also point out that the hydrodynamic picture of quantum mechanics originates in the seminal work of E. Madelung \cite{MA}, who interpreted $\rho^\e, J^\e$ as 
a description of a continuum fluid instead of a single particle.
\end{remark}
Identifying the current as $J^\e = \rho^\e u^\e$ we can formally derive from \eqref{qhd} the following equation for $u^\e$:
\begin{equation}\label{ueq}
\partial_t u^\e + (u^\e \cdot \nabla) u^\e +  \nabla V = \frac{\e^2}{2}   \nabla \left( \frac{\Delta \sqrt{\rho^\e}}{ \sqrt{\rho^\e}} \right).
\end{equation}
The right hand side can be seen as the gradient of the so-called \emph{Bohm potential} (or quantum potential), given by
\begin{equation}
\label{bohmpot}
V^\e_B:= -\frac{\e^2}{2} \frac{\Delta \sqrt{\rho^\e}}{ \sqrt{\rho^\e}}.
\end{equation}
Plugging \eqref{ueq} into \eqref{peq} we finally arrive at the following system of ordinary differential equations:
\begin{equation}
\label{bohm2}
\left \{
\begin{aligned}
& \,  \dot X^\e= P^\e , \\
& \, \dot P^\e = - \nabla V(X^\e ) -   \nabla V^\e_B(t, X^\e) ,
\end{aligned}
\right. 
\end{equation}
subject to the initial data $X^\e(0, x) = x$ and $P^\e(0, x) = u^\e(0,x)$, where $u^\e(0,x)$ is the initial velocity given by
$$
u_0^\e(x)=  \e \im \left(\frac{ \nabla \psi_0^\e( x)}{ \psi_0^\e(x)} \right ) .
$$
\begin{remark} Note that the system \eqref{bohm2} fully determines the quantum mechanical dynamics. 
It can be regarded as a system of ordinary differential equations, parametrized by the spatial variable $x\in \R^d$ through the initial data, 
where the position density $\rho^\e$ (and its derivatives up to order three) have to be determined additionally. 
At least numerically, $\rho^\e$ can be computed via ray tracing methods or particle methods \cite{DDP, HoEa, WyTr}, based on the push-forward formula \eqref{push}, i.e. 
evolving
$$
\rho^\e(t,x) \approx \sum_{n\in \N} c_n \, \delta (x - X^\e(t,x_n)).
$$
Strictly speaking, though, this requires the trajectories $X^\e(t,x)$ for all $x\in \R^d$ at time $t\in \R$. However, it approximately 
yields the solution of the continuity equation \eqref{conservation}, 
without solving the Schr\"odinger equation. 
This fact makes the Lagrangian reformulation interesting for numerical simulations, in particular in quantum chemistry, see e.g. \cite{BeMe, GMB, NeFre, WKH} for 
applications and \cite{WyTr} for a general overview. 
However, we caution that the Bohm potential $V_B$ is singular at nodes of the wave function, 
which generates significant numerical difficulties in actual computations (cf. \cite{DDP} for more details).
\end{remark}

In order to give \eqref{bohm2} a precise mathematical meaning we shall in the following introduce what we call 
\emph{Bohmian measures on phase space} $\R^d_x \times \R^d_p$. To this end we denote by $\mathcal M^+ (\R^d_x \times \R^d_p)$ the 
set of non-negative Borel measures on phase-space and by $\langle  \cdot, \cdot \rangle$ the corresponding 
duality bracket between $ \mathcal M (\R^d_x \times \R^d_p)$ and $C_0 (\R^d_x \times \R^d_p)$, where $C_0 (\R^d_x \times \R^d_p)$ 
is the closure (with respect to the uniform norm) of the set of continuous functions with compact support.

\begin{definition} \label{defbohm} 
Let $\e>0$ be a given scale and $\psi^\e \in H^1_\e(\R^d)$ be a sequence of wave functions with corresponding densities $\rho^\e $, $J^\e $. 
Then, the associated \emph{Bohmian measure} $\beta^\e \equiv \beta^\e[\psi^\e ]\in \mathcal M^+(\R^d_x \times \R^d_p)$ is given by
$$
\langle  \beta^\e , \varphi \rangle := \int_{\R^d} \rho^\e(x)  \varphi \left(x, \frac{J^\e(x)}{\rho^\e(x)} \right) dx , \quad \forall \, \varphi \in C_0(\R^d_x \times \R^d_p).
$$
\end{definition}
Note that in the definition of $\beta^\e$ a fixed scale $\e$ is imposed via the scaling of the gradient in the definition of the current density \eqref{densities}.
Formally, we shall denote the Bohmian measure by 
\begin{equation}\label{form}
\beta^\e(x,p) = \rho^\e(x)\, \delta \left( p-\frac{J^\e(x)}{\rho^\e(x)} \right) \equiv 
|\psi^\e(x)|^2\,\delta \left(p - \e  \im \left(\frac{ \nabla \psi^\e(x)}{ \psi^\e(x)} \right ) \right),
\end{equation} 
where $\delta$ is the usual delta distribution on $\R^d$. Obviously, \eqref{form} defines a continuous non-negative distribution on phase space. 
In addition, the first two moments of $\beta^\e$ satisfy
\begin{align*}
\int_{\R^d} \beta^\e(x, dp) = \rho^\e(x), \quad \int_{\R^d} p \, \beta^\e(x, dp) = \rho^\e(x) u^\e(x) \equiv J^\e(x).
\end{align*}
However, higher order moments of $\beta^\e$ in general \emph{do not} correspond quantum mechanical probability densities (defined via quadratic expressions of $\psi^\e$). 
In particular, the second moment of $\beta^\e$ yields
\begin{align*}
\int_{\R^{d}} \frac{|p|^2}{2} \, \beta^\e(x, dp) =   \frac{1}{2} \rho^\e(x)|u^\e(x)|^2.
\end{align*}
In classical kinetic theory, this would be interpreted as the kinetic energy density of the particle. However, in view of
\begin{equation}\label{Ekin}
E_{kin}[\psi^\e]: = \frac{\e^2}{2} \int _{\R^d} | \nabla \psi^\e (x) |^2 dx =\frac{1}{2}  \int _{\R^d} \frac{|J^\e(x)|^2}{\rho^\e(x)} +  \frac{\e^2}{2} \int _{\R^d}  | \nabla \sqrt{\rho^\e} |^2 dx ,
\end{equation}
we see that the second moment of $\beta^\e$ is not what in quantum mechanics would be called a kinetic energy density, 
since it does not account for the second term $\propto \e^2$. Note that this term formally goes to zero in the classical limit $\e \to 0_+$.

To proceed further, we shall introduce the following mapping on phase space,
\begin{equation}\label{flow}
\Phi^\e_t: (x, p) \mapsto (X^\e(t,x,p),  P^\e(t,x,p))
\end{equation}
where $X^\e, P^\e$ formally solve the ODE system \eqref{bohm2} for \emph{general initial data} $x,p\in \R^d$. Note that this $\Phi^\e_t$ is not necessarily well defined as a mapping on the whole phase space. 
However, from Proposition \ref{existence}, it is straightforward to conclude the following existence result.

\begin{lemma}\label{lemLag} Under the same assumptions as in Proposition \ref{existence}, the mapping $\Phi_t^\e$ exists globally in-time 
for almost all $(x,p) \in \R^{2d}$, relative to the measure 
$$\beta^\e_0(x,p)= \rho_{0} ^\e(x)\, \delta(p- u^\e_0(x)). $$ 
Moreover $\Phi_t^\e$ is continuous w.r.t. $t\in \R$  on its maximal open domain and 
\begin{equation*}
\beta^\e(t) = \Phi^\e_t \, \# \,  \beta_{0} ^\e.
\end{equation*}
\end{lemma}
\begin{proof} First note that $\Phi^\e_{t}$ when restricted to $\{ {\rm graph}(u^\e_0)\}\subset\R^d_x\times\R^d_p$ is well defined $\beta^\e_0-a.e.$, 
since the map $X_t^\e$ established in Proposition \ref{existence} does not run into nodes of $\psi^\e(t,\cdot)$ 
for almost all $x$ relative to $\rho^\e_0$.
Now, let us w.r.o.g. consider test function $\varphi(x,p)= \sigma(x) \chi(p)\in C_b(\R^{2d})$ and denote $u^\e= J^\e/\rho^\e$. Then, we have 
\begin{align*}
\langle  \beta^\e(t), \varphi \rangle = & \, \int_{\R^d} \sigma(x) \chi(u^\e(t,x)) \rho^\e(t,x)  dx\\
= & \, \int_{\R^d} \sigma(X^\e(t,x)) \chi(u^\e(t,X^\e(t,x))) \rho_0(x)dx ,
\end{align*}
where for the second equality we have used \eqref{push}. By definition $u^\e(t,X^\e(t,x)) = P^\e(t,x)$, hence
\begin{align*}
\langle  \beta^\e(t), \varphi \rangle =  \, \int_{\R^d} \sigma(X^\e(t,x)) \chi(P^\e(t,x)) \rho_0(x)dx  = 
\langle \Phi^\e_t \, \# \,  \rho_0^\e \, \delta_{p=P^\e(0)} , \sigma \chi \rangle .
\end{align*} 
Since $P^\e(0,x)  \equiv u^\e(0, X^\e(0,x))=u_0^\e(0,x)$ the assertion of the lemma is proved.
\end{proof}

Thus $\beta^\e(t)$ is supported on a subset of phase space given by the graph of the velocity field and evolves through the quantum 
mechanical trajectories, induced by \eqref{bohm2}, in the sense of a push-forward for measures. 
This result is conceptually important, since it established the existence of a $t$-parametrized family of phase space measures $\beta^\e(t)\equiv \beta^\e(t,x,p)$, 
which can readily be compared to the concept of Wigner functions $w^\e(t,x,p)$, see Section \ref{sec: Wig}, and which encodes the \emph{full} quantum mechanical dynamics of $\rho^\e(t,x)$ and $J^\e(t,x)$. 
It is therefore natural to consider the limit of $\beta^\e(t)$ as $\e \to 0_+$ in order to gain insight into the classical limit of Bohmian mechanics. 
This will be the main task of the upcoming sections.

\begin{remark} \label{kinremark} 
Lemma \ref{lemLag} formally allows to interpret $\beta^\e(t)$ as a solution of the following \emph{nonlinear kinetic equation}
\begin{equation}\label{kinetic}
\partial_t \beta^\e + p \cdot \nabla_x \beta^\e - \nabla_x (V +   V^\e_B) \cdot \nabla_p \beta^\e = 0,\quad \rho^\e(t,x) = \int_{\R^d} \beta^\e(t,x,dp),
\end{equation}
subject to initial data $\beta_0^\e$ as given in Lemma \ref{lemLag}. 
As before, equation \eqref{kinetic} can be seen as a conservation law in phase space (endowed with a complex structure). 
The main problem of \eqref{kinetic} is, that 
the term $\nabla_x V^\e_B \cdot \nabla_p \beta^\e$ can not be defined in the sense of distributions in a straightforward way. 
We remark, however, that in the purely diffusive setting of so-called \emph{quantum drift diffusion models}, this mathematical 
difficulty was overcome by using Wasserstein gradient-flow techniques \cite{GST}. We believe that a combination of \cite{GST} with 
the results given in \cite{AMGA, GNT} can lead to rigorous mathematical results on \eqref{kinetic}, opening the door for a new interpretation of Bohmian mechanics via optimal 
mass-transportation. 
From a mathematical point of view, the study of \eqref{kinetic} is also interesting for general measure valued initial data, even though,
the connection with the Schrödinger equation is lost in such a situation.
\end{remark}

\section{The classical limit of Bohmian measures} 

We recall that the assumptions on the initial wave function $\psi_0^\e$ together with the arguments given at the beginning of Section \ref{sec:Lag} imply that 
for all $t\in \R$ the solution of the Schrödinger equation 
$\psi^\e(t)$ is uniformly bounded in $H^1_\e(\R^d)$ as $\e \to 0_+$, with a bound independent of time (namely, the initial energy). 
Since the latter will be the main technical assumption needed from now on, we shall for the sake of notation 
suppress any time-dependence in the following  and formulate results on Bohmian and Wigner measures associated to general sequences of $L^2$ functions $\psi$ with uniformly (in $\e$) bounded 
mass and energy. In Section 6 we shall get back to time-dependent setting and connect it to the results of the previous sections. 

\subsection{Existence of limiting measures} We start with the following basic lemma, which ensures existence of a classical limit of  $\beta^\e$.

\begin{lemma} \label{convlem1} 
Let $\psi^\e$ be uniformly bounded in $L^2(\R^d)$. 
Then, up to extraction of sub-sequences, there exists a limiting measure $\beta^0 \equiv \beta \in \mathcal M^+(\R^d_x \times \R^d_p)$, such that
$$
\beta^\e  \stackrel{\e\rightarrow 0_+
}{\longrightarrow} \beta \quad \text{in $\mathcal M^+(\R^d_x \times \R^d_p) \, {\rm w}-\ast$}.
$$
\end{lemma}
\begin{proof}
In view of Definition \ref{defbohm} we have, for all test functions $\varphi \in C_0(\R^d_x \times \R^d_p)$:
$$
|\langle  \beta^\e, \varphi \rangle| \le  \| \varphi \|_{L^\infty(\R^{2d})}  \| \rho^\e  \|_{L^1(\R^d)} <+ \infty,
$$
uniformly in $\e$, by assumption. By compactness, we conclude that there exists a sub-sequence $\{\e_n\}_{n \in \N} $, tending to zero as $n\to \infty$, such that 
$\beta^{\e_n} \stackrel{n \rightarrow  \infty}{\longrightarrow} \beta $
in $ \mathcal M^+(\R^d_x \times \R^d_p)$ weak -- $\ast$.
\end{proof}
When $\psi^\e = \psi^\e (t)$ evolves according to the Schrödinger equation with initial data such that $\| \psi^\e_0 \|_{L^2} = 1$, then $\beta^\e$ is in $L^\infty (\R_t, \mathcal M^+ (\R^d_x\times \R^d_p))$ 
uniformly as $\e \to 0_+$, since $\rho^\e$ is in $L^\infty (\R_t, L^1(\R_x ^d))$ uniformly as $\e \to 0_+$. Thus there exists a sub-sequence (which we shall denote by the same symbol) and a $t$-parametrized 
family of limiting probability measures $\beta^0 = \beta^0 (t)$, such that $\beta^\e$ tends to $\beta^0$ in $L^\infty (\R_t, \mathcal M^+ (\R^d_x\times \R^d_p))$ weak -- $\ast$.
This is of importance when we shall get back to Schrödinger wave functions in Section 6, in particular for Proposition 6.1.

Next, we shall be concerned with the classical limits of the densities $\rho^\e, J^\e$. Since they are both uniformly bounded in $L^1(\R^d)$, provided $\psi^\e$ is uniformly bounded in $H^1_\e(\R^d)$, 
we conclude, that, up to extraction of a subsequence, it holds
\begin{equation}\label{limdensities}
\rho^\e  \stackrel{\e\rightarrow 0_+
}{\longrightarrow} \rho , \  \ \text{in $ \mathcal M^+(\R^d_x;\R) \, {\rm w}-\ast$,} \quad J^\e  \stackrel{\e\rightarrow 0_+}{\longrightarrow} J, \ \text{in $ \mathcal M^+(\R^d_x;\R^d) \, {\rm w}-\ast$}.
\end{equation}
Moreover, it has been proved in \cite{GaMa}, that $J \ll \rho$ in the sense of measures and thus, by the Radon-Nikodym theorem there exists a measurable function 
$u$, such that
\begin{equation}\label{u}
dJ = u(x) d\rho. 
\end{equation}
Formally, the function $u(x)\in \R^d$ can be interpreted as the classical limit of the Bohmian velocity field $u^\e$.
The following statement gives the connection between the limits $(\rho, J)$ and $\beta$.

\begin{lemma} 
Let $\psi^\e$ be uniformly bounded in $H^1_\e(\R^d)$. Then  
\begin{equation}\label{limdens}
\rho(x)= \int_{\R^d} \beta(x, dp) , \quad 
J (x)= \int_{\R^d} p \, \beta(x, dp) .
\end{equation}
Moreover, we also have
\begin{equation}\label{limmass}
\lim_{\e \to 0_+} \iint_{\R^{2d}} \beta^\e[\psi^\e] (dx, dp) = \iint_{\R^{2d}} \beta(dx, dp) ,
\end{equation}
provided that the sequence $\psi^\e$ is compact at infinity, i.e.
$$
\lim_{R\to \infty} \lim_{\e \to 0_+} \int_{|x|\ge R}|\psi^\e(x)|^2\, dx = 0 . 
$$
\end{lemma}

Thus, the classical limit of the densities $\rho^\e, J^\e$ can be obtained from the limiting Bohmian (phase space) measure 
$\beta$ by taking the zeroth and first moment. In addition no mass is lost during the limiting process at $|x| + |p| = + \infty$, if in addition $\psi^\e$ is compact at infinity.

\begin{remark} Note that the property of $\psi^\e$ being compact at infinity \cite{GMMP, LiPa} can be rephrased as $|\psi^\e|^2$ being \emph{tight}.
\end{remark}

\begin{proof} We first prove assertion \eqref{limdens} for $\rho^\e$. To this end let $ \sigma \in C_0(\R^d)$ and write
\begin{align*}
\iint_{\R^{2d}} \sigma(x)  \beta^\e(dx, dp)  = & \ \iint_{\R^{2d}}\sigma(x) \chi_R(p) \beta^\e(dx, dp)\\
& \  +\iint_{\R^{2d}} \sigma(x)  (1-\chi_R(p)) \beta^\e(dx, dp) ,
\end{align*}
where for a given cut-off $R>0$, $\chi_R\in C_0(\R^d)$, such that: $0\le \chi_R\le 1$ and $\chi_R(p)=1$ for $|p|< R$, as well as $\chi(p) = 0$ for $|p|>R+1$.
In view of Lemma \ref{convlem1} the first integral on the r.h.s. converges 
$$
\iint_{\R^{2d}}\sigma(x) \chi_R(p) \beta^\e(dx, dp) \stackrel{\e\rightarrow 0_+}{\longrightarrow}\iint_{\R^{2d}}\sigma(x) \chi_R(p) \beta(dx, dp).
$$
On the other hand, the second integral on the r.h.s. is 
$$
\iint_{\R^{2d}} \sigma(x)  (1-\chi_R(p)) \beta^\e(dx, dp) \le \int_{\R^{d}} \rho^\e(x) \sigma(x) {\bf 1}_{\{ |u^\e| > R\} } \, dx,
$$
where $u^\e= \frac{J^\e}{\rho^\e}$ and ${\bf 1}_{\Omega}$ denotes the indicator function of a given set $\Omega\subseteq \R^d$. We can now estimate 
$$
\int_{\R^{d}} \rho^\e(x) \sigma(x) {\bf 1}_{\{ |u^\e| > R\} } \, dx \le \frac{1}{R}  \int_{\R^{d}} |J^\e(t,x)| \sigma(x) \, dx \le \frac{C}{R} ,
$$
where $C\in \R_+$ is independent of $\e$. Here, the last inequality follows from \eqref{Jest} together with the uniform bound of $\psi^\e$ in $H^1_\e(\R^d)$. We can therefore 
take the respective limits $\e\to 0_+$ and $R \to + \infty$, to obtain the desired statement for the position density $\rho^\e$. The assertion \eqref{limdens} for 
$J^\e$ can be shown analogously. Finally, in order to prove  \eqref{limmass}, we refer to \cite{GMMP, LiPa}, where it is shown that 
$$
\lim_{\e \to 0_+} \int_{\R^d}\rho^\e(x) \, dx = \int_{\R^d}\rho(x) \, dx , 
$$
provided $\rho^\e= |\psi^\e|^2$ is tight. Jointly with \eqref{limdens}, this directly implies \eqref{limmass}.
\end{proof}

The main task of this work is henceforth to study the limit $\beta\in \mathcal M^+(\R^d_x\times\R^d_p)$. In particular, we want to understand under which circumstances $\beta$ is mono-kinetic.
\begin{definition}
We say that $\beta \in \mathcal M^+(\R^d_x\times\R^d_p)$ is \emph{mono-kinetic}, if there exists a measure $\rho \in \mathcal M^+(\R_x^d)$ and a function $u$ defined $\rho - a.e.$, such that 
\begin{equation}\label{mono}
\beta(x,p) = \rho(x) \, \delta (p-u(x)).
\end{equation}
\end{definition}
Obviously, for every fixed $\e >0$ the Bohmian measure $\beta^\e$ is mono-kinetic by definition, see \eqref{form}. Note however, that the limit statements for $\rho^\e$ and $J^\e$ given in \eqref{limdensities}, 
do \emph{not} allow us to directly pass to the limit in $\beta^\e$. Thus, in general we can not expect the limiting Bohmian measure $\beta$ to be of the form \eqref{mono}.  
In order to obtain further insight into the situation, we shall establish in the upcoming subsection a connection between $\beta$ and the, by now classical, theory of Young measures. 

\subsection{Connection between Bohmian measures and Young measures} 

Consider a sequence $f_\e: \R^d\to \R^m$ of measurable functions. Then, we recall that  
there exists a mapping $\mu_x\equiv \mu(x): \R^d \to \mathcal M^+(\R^m)$, called the \emph{Young measure associated to the sequence $f_\e$}, 
such that $x\mapsto \langle \mu(x), g \rangle$ is measurable for all $g \in C_0(\R^m)$,
and (after selection of an appropriate subsequence):
$$
\lim_{\e \to 0} \int_{\R^d} \sigma ( x, f_\e(x)) \, dx = \int_{\R^d} \int_{\R^m} \sigma ( x, \lambda) d\mu_x(\lambda) \, dx,
$$
for any function $\sigma \in L^1(\Omega; C_0(\R^m))$, cf. \cite{Ba, Pe1, Pe2}. In view of Definition \ref{defbohm} we expect a close connection between the classical limit of 
Bohmian measures and Young measures. To this end, we shall first state one of the key technical lemmas of this work.

\begin{lemma} Let $\psi^\e$ be uniformly bounded in $H^1_\e(\R^d)$ with corresponding densities $\rho^\e, J^\e\in L^1$. 
Then, for $x\in \R^d$ a.e., there exists a Young measure $$\mu_x: \R^d_x \to \mathcal M^+(\R_r \times \R_\xi^{d}),$$ associated to the 
pair $(\rho^\e, J^\e)$, such that 
\begin{equation}\label{young}
\beta (x,p) \ge \int_0^\infty r^{d+1} \mu_x (r, r p) \, dr,
\end{equation}
in the sense of measures, 
with equality if $\rho^\e  \rightharpoonup  \rho$ in $L^1(\R^d)$, as $\e\rightarrow 0_+$. In the latter case, $\mu_x$ is a probability measure on $\R^{d+1}$.
\end{lemma}
The property of weak convergence of the particle density is crucial in order to express the limiting Bohmian measure $\beta$ by (a moment of) the Young measure 
associated to $\rho^\e, J^\e$.

\begin{proof} Assume weak convergence of $\rho^\e  \rightharpoonup  \rho$ in $L^1(\R^d)$, as $\e\to 0_+$. Thus, by the Dunford-Pettis theorem $\rho^\e$ is uniformly integrable. 
Next, consider the sequence 
$$
\alpha^\e(x):= \rho^\e(x)  \varphi \left(x, \frac{J^\e(x)}{\rho^\e(x)} \right) ,  
$$
for $\varphi \in C_0^\infty(\R^{2d}; \R)$ such that w.r.o.g. $\| \varphi \|_{L^\infty} = 1$. 
Then $|\alpha^\e| \le \rho^\e$, and in addition the sequence $\alpha^\e$ is uniformly integrable. Thus 
$\alpha^\e  \rightharpoonup \alpha^0$ in $L^1(\R^d)$ weakly, even though $J^\e$ does not necessarily converge weakly in $L^1$. 
In view of Definition \ref{defbohm}, we obviously have  $\langle  \beta^\e , \varphi \rangle = \int \alpha^\e(x) dx$, 
and thus also in the limit $\langle  \beta , \varphi \rangle = \int \alpha^0(x) dx$.

On the other hand, we know that for $x \in \R^d$, the mapping
$$ 
\alpha: (x, r, \xi) \to r \,  \varphi \left(x, \frac{\xi}{r} \right)
$$
is continuous in $r, \xi$ and measurable in $x$, hence a Carath\'eodory function (cf. \cite{Pe1}). From what we have seen before, we know 
that $\alpha(x, \rho^\e(x), J^\e(x)) \equiv \alpha^\e(x)$ converges weakly in $L^1(\R^d, \R)$ and thus Theorem 2.3 in \cite{Pe1} asserts the existence of a probability measure $\mu_x$, 
associated to $(\rho^\e, J^\e)$, such that
\begin{align*}
\lim_{\e \to 0} \int_{\R^d} \alpha (x, \rho^\e(x), J^\e(x)) \, dx = & \, \int_{\R^d} \int_{\R^{d+1}} r \varphi \left(x, \frac{\xi}{r} \right) d\mu_x(r, \xi) \, dx\\
= & \, \int_{\R^d} \int_{\R^{d+1}} r^{d+1} \varphi (x, p) d\mu_x(r, r p ) \, dx,
\end{align*}
where the second line follows from the simple change of variables $ r p = \xi $. Since $\rho^\e\ge 0$ the Young measure $\mu_x$ has to be 
supported in $[0,\infty)\times \R^d$ and thus, we obtain
$$
\langle  \beta , \varphi \rangle =  \int_{\R^d} \int_{[0,\infty)\times \R^d} r^{d+1} \varphi (x, p) d\mu_x(r, r p ) \, dx,
$$
i.e. the assertion of the theorem, provided $\rho^\e  \rightharpoonup  \rho$ in $L^1(\R^d)$. 

On the other hand, if we discard the assumption of weak $L^1$ convergence of $\rho^\e$, we infer the existence of a Young measure $\mu_x$, such that $\mu_x(\R^{d+1}) \le 1$, i.e. 
not necessarily a probability measure, and that (see also Proposition 4.4 in \cite{Pe1}):
$$
\liminf_{\e \to 0} \int_{\R^d} \alpha (x, \rho^\e(x), J^\e(x)) \, dx \ge  \int_{\R^d} \int_{[0,\infty)\times \R^d} r^{d+1} \varphi (x, p) d\mu_x(r, r p ) \, dx,
$$
This concludes the proof.
\end{proof}
To proceed further we recall the following definition: A sequence of (measurable) functions $\{ f_\e\}_{0<\e\le 1}: \R^d\to \R$ is said to \emph{converge in measure} to (the function) $\widetilde f $ as $\e \to 0$, if for every $\delta>0$:
$$
\lim_{\e\to 0} {\rm meas}\big(\{ | f_\e (x) - \widetilde f(x)| \ge \delta \} \big) = 0.
$$
Note that if in addition $0 \le f_\e   \stackrel{\e\rightarrow 0_+}{\longrightarrow} f$ in $\mathcal M^+(\R^d)$ ${\rm w}-\ast$, then in general: $\widetilde f \le f$ in the sense of measures.

\begin{theorem} \label{thyoung1} Let $\psi^\e$ be uniformly bounded in $H^1_\e(\R^d)$ with corresponding densities $\rho^\e, J^\e\in L^1$. If $\rho^\e  \stackrel{\e\rightarrow 0_+
}{\longrightarrow}  \rho$ in $L^1(\R^d)$ strongly and $J^\e  \stackrel{\e\rightarrow 0_+
}{\longrightarrow} \widetilde J$ in measure, then 
$\beta$ is mono-kinetic, i.e.
\begin{equation}\label{mono1}
\beta(x, p) =  \rho(x) \, \delta \left(p - \frac{\widetilde J(x)}{ \rho(x)}\right).
\end{equation}
and in addition $\widetilde J = J$, where $J$ is the measure weak $-*$ limit established in \eqref{limdensities}.
\end{theorem}

\begin{proof} We first note that strong convergence of $\rho^\e$ in $L^1(\R^d)$ implies that $\rho^\e  \stackrel{\e\rightarrow 0_+}{\longrightarrow}  \rho$ in measure. 
Since it is known that convergence in measure of $\rho^\e, J^\e$ is equivalent to the fact that $\mu_x$ is only supported in a single point of $\R^{d+1}$, cf. \cite[Proposition 4.3]{Pe1}, we conclude 
\begin{equation}\label{monoyoung}
\mu_x(r, \xi) = \delta (r-  \rho(x)) \, \delta (\xi - \widetilde J(x)).
\end{equation} 
In addition, since strong convergence of $\rho^\e$ in $L^1(\R^d)$ also implies weak convergence, we can insert \eqref{monoyoung} 
into \eqref{young} (with equality), to obtain
$$
\beta (x, p) = \rho^{d+1}(x) \, \delta ( \rho (x) p- \widetilde J(x)) ,
$$
and a simple change of variable yields \eqref{mono1}. By computing the first moment of \eqref{mono1} w.r.t. $p$ and keeping in mind \eqref{limdens},
we conclude that $\widetilde J = J$ in this case.
\end{proof}
Recalling the results of \cite{GaMa}, we infer that the limiting measure $\beta$ given by \eqref{mono1} can be rewritten as
\begin{equation*}
\beta(x, p) =  \rho(x) \, \delta \left(p -u(x)\right),
\end{equation*}
where $u$ is defined $\rho - a.e.$ by \eqref{u}. In this case $u$ can be considered the classical limit of the Bohmian velocity field.
\begin{remark} In the case where $\rho^\e  \stackrel{\e\rightarrow 0_+}{\longrightarrow} \widetilde  \rho$ in measure, but not necessarily weakly in $L^1(\R^d)$, we still 
know that the Young measure is given by \eqref{monoyoung}. However, in such a situation, we can only conclude that
$$
\beta(x,p) \ge \widetilde \rho(x) \, \delta \left(p - \frac{\widetilde J(x)}{ \widetilde \rho(x)}\right).
$$
If $\widetilde \rho = 0$, which can happen in principle, no information on $\beta$ is provided.
\end{remark}

\subsection{An alternative point of view}

One might want to describe the classical limit of $\beta^\e$ by the Young measure $\nu_x$ associated to $(\rho^\e, u^\e)$ instead of the one associated to $(\rho^\e, J^\e)$. 
However, the problem with using $\nu_x$ instead of $\mu_x$ is the fact that $u^\e:=\frac{J^\e}{\rho^\e}$ is only defined $\rho^\e-a.e.$ and thus, 
we can not directly obtain a result for $\nu_x$ analogous to the one given in Theorem \ref{thyoung1}. Rather, we need to assume the existence of an 
appropriate \emph{extension} of $u^\e$ defined on all of $\R^d$, which satisfies the required convergence in measure. In this case, a change of variables yields
$$
\nu_x(r,\xi) = r^d \mu_x( r, r \xi), \quad (r, \xi) \in \R^{1+d}.
$$
Thus, instead of \eqref{young} we obtain 
\begin{equation}
\label{altyoung}
\beta (x,p) \ge \int_0^\infty r \nu_x (r, p) \, dr .
\end{equation}
Despite the above mentioned drawback, the measure $\nu_x$ is still useful to show that the \emph{converse statement} of Theorem \ref{thyoung1} \emph{is not true in general}. 
To this end, we assume $\beta$ to be mono-kinetic, i.e. 
$\beta(x, p) =  \rho(x) \, \delta \left(p -u(x)\right)$, from which we conclude from \eqref{altyoung} that 
\begin{align*}
\text{supp}_{x,p} \left( \int_0^\infty r \nu_x (r, p) \, dr  \right) \subseteq \overline{\big\{ (x,p)\in \R^{2d}: p = u(x) , x \in \text{supp $\rho$}  \big \}}.
\end{align*}
Thus
\begin{align*}
\text{supp}_{r,p} \nu_x \subseteq \overline{\big \{ (r,p)\in \R^{2d}: p = u(x) , r >0 \}} \cup \{ (r=0,p): p\in \R^{d} \big \}
\end{align*}
and we consequently infer
$$
\nu_x(r,p) = \omega_x(r) \, \delta(p - u(x)) + \delta (r)  \gamma_x(p),
$$
where $\text{supp}_{r} \, \omega_x \subseteq (0,\infty)$. The appearance of the second term on the right hand side makes the converse statement of Theorem \ref{thyoung1} fail in general. In other words, the 
fact that $\beta$ is mono-kinetic does not imply that $\nu_x$ is a delta distribution in $p$, which makes it impossible to conclude the strong convergence of $u^\e$ (or $J^\e$). This fact can be further illustrated by 
the following example.

\begin{example} For any $\psi^\e\in L^2(\R^d)$ we can write 
\begin{equation}\label{reppsi}
\psi^\e(x) = \sqrt{\rho^\e(x)} e^{i S^\e(x)/ \e},
\end{equation}
where $S^\e(x)\in \R$ is defined $\rho^\e - a.e.$, up to additive integer multiples of $2\pi$. 
In this representation (which should not be confused with the WKB ansatz to be discussed in Section \ref{sec:WKB}) the current density reads $J^\e = \rho^\e \nabla S^\e$. 
Assume now, that for some measurable set $\Omega\subset \R^d$ we have
$$
\rho^\e = \rho^\e_1 {\bf 1}_{\Omega} + \rho^\e_2 {\bf 1}_{\{\R^d / \Omega\}}
$$
with $\rho_1^\e  \stackrel{\e\rightarrow 0_+}{\longrightarrow} 0$ in $L^1(\Omega)$ strongly and $\rho_2^\e  \stackrel{\e\rightarrow 0_+}{\longrightarrow} \rho_2\not =0$ in $L^1(\R^d / \Omega)$ weakly. 
Similarly, we assume $$
S^\e = S^\e_1 {\bf 1}_{\Omega} + S^\e_2 {\bf 1}_{\{\R^d / \Omega\}}
$$
with $\nabla S_1^\e  \stackrel{\e\rightarrow 0_+}{\longrightarrow} \nabla S_1$ in $L^\infty(\Omega)$ weak$-\ast$ (but not strongly) 
and $\nabla S_2^\e  \stackrel{\e\rightarrow 0_+}{\longrightarrow} \nabla S_2$ almost everywhere on $\R^d / \Omega$. 
Then, one easily checks that 
$$
\beta(x,p) = \rho_2 {\bf 1}_{\{\R^d / \Omega\}} \, \delta (p- \nabla S_2(x)),
$$
i.e. mono-kinetic. The corresponding Young measure however, is found to be
$$
\nu_x(r,p) = \omega_x(r){\bf 1}_{\{\R^d / \Omega\}} \, \delta(p - \nabla S_2(x)) + \delta (r){\bf 1}_{ \Omega }  \gamma_x(p),
$$
where $\omega_x(r)$ is the Young measure of $\rho^\e_2 {\bf 1}_{\{\R^d / \Omega\}}$ and $\gamma_x(p)$ is the Young measure of $\nabla S_1^\e{\bf 1}_{ \Omega}$. 
In other words, the oscillations within $S^\e_1$ do not show in the limiting Bohmian measure $\beta$ (since the corresponding limiting density vanishes), 
but they \emph{do occur} in the corresponding Young measure.
\end{example}

\begin{remark} \label{rem: measXP} It is certainly interesting to see whether the analysis given above directly yields information on the classical limit of the Bohmian trajectories $X^\e$, $P^\e$ defined by \eqref{bohm2}. 
To this end, let $$ \Upsilon_{t,x}: \R_t \times \R^d_x \to \mathcal M^+(\R^d_y \times \R^d_p)$$ be the Young measure associated to the Bohmian trajectories  $(X^\e(t,x), P^\e(t,x))$ and assume for simplicity that 
$\rho^\e_0  \stackrel{\e\rightarrow 0_+}{\longrightarrow}  \rho_0$ in $L^1(\R^d)$ strongly. Then we conclude from the proof of Lemma \ref{lemLag} that for all test-functions 
$\varphi \in C_0(\R^d_x\times\R^d_p)$, $\chi\in C_0(\R_t)$ it holds
$$
\int_{\R} \chi  (t) \iint_{\R^{2d}} \varphi(x,p) \beta^\e (t, dx, dp) dt = \int_{\R} \chi  (t) \int_{\R^{d}} \varphi(X^\e(t,x),P^\e(t,x))\rho_0^\e(x) \, dx \, dt .
$$
Passing to the limit $\e \to 0_+$ on both sides we find that 
\begin{equation}\label{measXP}
\beta(t,y,p) = \int_{\R^d}  \Upsilon_{t,x} (y,p )  \rho_0(x) dx .
\end{equation}
Formula \eqref{measXP} implies that $\beta(t)$ can be uniquely determined from $\Upsilon_{t,x}$ but in general not the other way around. An immediate conclusion of \eqref{measXP} is 
that a.e. in $t\in \R: \, (y,p) \in \text{supp} \, \beta(t)  $ if and only if there exists an $x \in \text{supp}\,  \rho_0$ such that $(y,p) \in \text{supp}\,  \Upsilon_{x,t}$. In addition we infer that if 
$$
X^\e  \stackrel{\e\rightarrow 0_+}{\longrightarrow}  X,\quad  P^\e  \stackrel{\e\rightarrow 0_+}{\longrightarrow}  P,
$$
in measure, i.e. $\Upsilon_{t,x}$ is only supported in a single point, then $\beta(t)$ is mono-kinetic. Conversely, though, from the fact that $\beta(t)$ is mono-kinetic we can only conclude directly that 
$$
\Upsilon_{t,x} = \upsilon_{t,x} \, \delta (p-P(t,x))
$$
where $\upsilon_{t,x}=\upsilon_{t,x}(y)$ is the Young measure associated to the sequence $X^\e(t,x)$. We shall use \eqref{measXP} 
as the basis for further investigations of the classical limit of Bohmian trajectories in a future work.  
Note, however, that the Young measure $\Upsilon_{t,x}$ carries more information than is needed in order to determine the classical limit of Bohmian trajectories.
\end{remark}

\section{Comparison to Wigner measures}\label{sec: Wig} 

In this section we shall compare the concept of Bohmian measures (and in particular their classical limit) to the 
well known theory of semi-classical measures, also called Wigner measures, see e.g. \cite{Ge, GMMP, LiPa} for a broader introduction. 
In the following, we denote the Fourier transform of a function $\varphi(x)$ by 
$$
\widehat \varphi(\xi):= \int_{\R^d} \varphi(x) e^{- i x\cdot \xi} dx.
$$

\subsection{Short review of Wigner measures} 

In order to obtain a phase space picture of quantum mechanics one usually considers the \emph{Wigner function} (or Wigner transformation)
$w^\e \equiv w^\e[\psi^\e]$, as introduced in \cite{Wi}:
\begin{equation}\label{wig}
w^\e (x,p): = \frac{1}{(2\pi)^d} \int_{\R^d}
\psi^\e\left(x-\frac{\e}{2}y 
\right)\overline{\psi^\e} \left(x+\frac{\e}{2}y
\right)e^{i y \cdot p}\,  dy. 
\end{equation}
In view of this definition, the Fourier transform of $w^\e$ w.r.t. $p$ is given by
\begin{align}\label{p-Fourier}
\widehat w^\e(x, y ) \equiv \int_{\R^d} w(x,p) e^{- i y \cdot p} dp = \psi^\e \Big (x+
\frac{\e}{2} y\Big) \overline{\psi^\e} \Big(x- \frac{\e}{2} y\Big),
\end{align}
and thus Plancherel's theorem together with a simple change of variables yields
$$
\|w^\e  \|_{L^2(\R^{2d}) }= \e^{-d} (2 \pi)^{-d/2} \|\psi^\e  \|^2_{L^2(\R^{d})} .
$$
The real-valued function $w^\e(t,x,p) $ acts as a quantum mechanical analogue for classical phase-space distributions. 
In particular, its moments satisfy
\begin{equation}\label{moments}
\rho^\e(x)= \int_{\R^d} w^\e (x,p) dp , \quad J^\e(x) =  \int_{\R^d} p \, w^\e (x,p) dp,
\end{equation}
where the integrals on the r.h.s. have to be understood in an appropriate sense, since $w^\e \not \in L^1(\R^d_x\times \R^d_p)$ in general. 
\begin{remark} More precisely, it is proved in \cite{LiPa, GMMP} that 
the Fourier transform of $w^\e$ w.r.t. $p$ satisfies $\widehat w^\e \in C_0(\R^d_y; L^1(\R^d_x))$ and likewise for the Fourier transformation of $w^\e$ w.r.t. $x\in \R^d$. 
This allows to define the integral of $w^\e$ via a limiting process after convolving $w^\e$ with Gaussians, cf. \cite{LiPa} for more details.
\end{remark}
The evolution equation for $w^\e(t,x,p) \equiv w^\e[\psi^\e(t)]$ is easily derived from the linear Schr\"odinger equation \eqref{schro}. It reads 
\begin{equation} \label{wignereq}
\partial _t w^\varepsilon+p \cdot \nabla_x w^\varepsilon + \Theta ^\varepsilon[V]w^\varepsilon =0, \quad w^\e(0, x,p) = w_0^\e(x,p),
\end{equation}
where $w_0^\e\equiv w^\e[\psi_0^\e]$ and  $\Theta^\varepsilon  [V]$ is a pseudo-differential operator
\begin{align*}
(\Theta^\e [V] f)(x,p) := -\frac{i}{(2\pi )^d } \iint_{\mathbb R^{2d}} \delta V^\e (x,y)   f (x,q)\ e^{i y \cdot (p- q)} \, dy \, dq ,
\end{align*}
with symbol $\delta V^\e$ given by
\begin{align*}\label{delta}
\delta V^\e (x,y)= \frac{1}{\e} \left( V\Big (x+
\frac{\e}{2} y\Big)-V\Big(x- \frac{\e}{2} y\Big)\right).
\end{align*}
Obviously, under the assumption \eqref{assumptionV} it holds $\delta V^\e \stackrel{\e\rightarrow 0_+
}{\longrightarrow} y \cdot \nabla_x V$, in which case \eqref{wignereq} formally simplifies to the classical \emph{Liouville equation} on phase space.

Note that the Wigner picture of quantum mechanics is \emph{completely equivalent} to the Schr\"odinger picture.
The main drawback of using $w^\e$ is that in general it can also take negative values and hence can not be regarded as a probability distribution. 
Nevertheless it has the following important property (see e.g. \cite{GMMP}): For any operator $\text{Op}^\e(a)$, defined by \emph{Weyl-quantization} of the corresponding classical symbol 
$a(x, p) \in \mathcal S(\R^d_x\times \R^d_p)$, one can compute the expectation value of $\text{Op}^\e(a)$ in the state $\psi^\e$ via
\begin{equation}\label{expect}
\langle \psi^\e, \text{Op}^\e(a)\psi^\e\rangle_{L^2} =  \iint_{\R^{2d}} a(x,p) w^\e(x,p) dx \, dp,
\end{equation}
where the right hand side resembles the usual formula from classical statistical mechanics. 
To proceed further, we recall the main result proved in \cite{LiPa, GMMP}:

\begin{proposition} Let $\psi^\e$ be uniformly bounded in $L^2(\R^d)$. Then, the set of Wigner functions $\{w^\e\}_{0<\e\le 1} \subset \mathcal S'(\R^d_x \times \R^d_p)$ is weak$-\ast$ compact 
and thus, up to extraction of subsequences
$$
w^\e[\psi^\e ]  \stackrel{\e\rightarrow 0_+
}{\longrightarrow} w^0\equiv w \quad \text{in $\mathcal S'(\R^d_x \times \R^d_p) \, {\rm w}-\ast$} ,
$$
where the limit $w\in \mathcal M^+(\R^d_x \times \R^d_p)$ is called the Wigner measure. If, in addition $\psi^\e \in H^1_\e(\R^d)$ uniformly, then we also have 
$$
\rho^\e(x)  \stackrel{\e\rightarrow 0_+
}{\longrightarrow} \rho (x)= \int_{\R^d} w(x, dp) , \quad 
J^\e(x)  \stackrel{\e\rightarrow 0_+
}{\longrightarrow} J (x)= \int_{\R^d} p \, w( x, dp) .
$$
\end{proposition}
This result allows us to exchange limit and integration on the limit on the right hand side of \eqref{expect} to obtain
\begin{equation*}
\langle \psi^\e, \text{Op}^\e(a)\psi^\e \rangle_{L^2} \stackrel{\e\rightarrow 0_+
}{\longrightarrow}  \iint_{\R^{2d}} a(x,p) w(x,p) \, dx \, dp.
\end{equation*}
The Wigner transformation and its associated Wigner measure therefore are highly useful tools to compute the classical limit of the expectation values of physical observables. 

In addition it is proved in  \cite{LiPa, GMMP}, that 
$
w(t)= \Phi_t  \, \# \,  w_{0} $, where $w_0$ is the initial Wigner measure and $\Phi(t)$ is the classical phase space flow given by the Hamiltonian ODEs
\begin{equation}
\label{hamode}
\left \{
\begin{aligned}
& \,  \dot X = P , \quad X(0,x,p)= x, \\
& \, \dot P= - \nabla V(X), \quad P(0,x,p) = p.
\end{aligned}
\right. 
\end{equation}
In other words $w(t)$ can be considered a weak solution of the Liouville equation. 
Note that $\Phi_t$ is formally obtained from \eqref{flow} in the limit $\e \to 0_+$. 
It is therefore natural to compare the Wigner measure associated $\psi^\e$ with the 
corresponding classical limit of the Bohmian measure associated to $\psi^\e$.

\subsection{The sub-critical case}  As a first step we shall prove the following basic result, relating $\beta$ and $w$ in the \emph{sub-critical case} w.r.t. to the scale $\e$.

\begin{theorem} \label{thsub}
Assume that $\psi^\e$ is uniformly bounded in $L^2(\R^d)$ and that in addition
\begin{equation}\label{consub}
\e \nabla \psi^\e   \stackrel{\e\rightarrow 0_+}{\longrightarrow} 0 , \quad \text{\rm in $L^2_{\rm loc} (\R^d)$.}
\end{equation}
Then, up to extraction of subsequences, it holds
$$
w(x,p) = \beta(x,p) \equiv \rho(x) \, \delta(p).
$$
\end{theorem}
This result can be interpreted as follows: Sequences of functions $\psi^\e$ which \emph{neither oscillate nor concentrate} on the scale $\e$ (but maybe on some larger scale), 
yield in the classical limit the \emph{same} mono-kinetic Bohmian or Wigner measure with $p=0$. Clearly, condition \eqref{consub} is 
propagated in time by the (semi-classically scaled) free Schrödinger group $U^\e(t) = e^{-it \Delta /(2\e)}$.
\begin{proof}
Let $\varphi \in C^\infty_0(\R^d_x \times \R^d_p)$ and write
\begin{align*}
\langle  \beta^\e, \varphi \rangle - \langle  \rho^\e \, \delta_{p=0}, \varphi \rangle = & \, \int_{\R^d} \rho^\e(x)  \varphi \left(x, \frac{J^\e(x)}{\rho^\e(x)} \right) dx - \int_{\R^d} \rho^\e(x)  \varphi \left(x, 0 \right) dx \\
= & \  \int_{\Omega}  \nabla_p \varphi \left(x, \eta^\e \right) \cdot J^\e(x) \ dx ,
\end{align*}
by using the mean value theorem. Using the fact that $\varphi \in C^\infty_0(\R^d_x \times \R^d_p)$, we can estimate
$$
|\langle  \beta^\e, \varphi \rangle - \langle  \rho^\e \, \delta_{p=0}, \varphi \rangle| \le C \int_{\Omega}  | J^\e(x)| \ dx \le C \|  \psi^\e \|_{L^2(\Omega)} \| \e \nabla \psi^\e \|_{L^2(\Omega)},
$$
and since, by assumption, $  \e \nabla \psi^\e \to 0$ in $L^2_{\rm loc} (\R^d)$ we conclude that the limiting Bohmian measure is of the form given above. 

In order to prove the same statement for the Wigner function we again use the mean value theorem to write
\begin{equation}\label{middle}
\psi^\e \Big (x\pm \frac{\e}{2} y\Big) = \psi^\e (x) \pm \frac{\e}{2} \int_{0}^1 \nabla \psi^\e \Big (x\pm \frac{\e s}{2}  y \Big ) \cdot y \, ds,
\end{equation}
and consider the Fourier transformation of $w^\e$ w.r.t. the variable $p\in \R^d$, i.e.
\begin{align*}
\widehat w^\e(x, y ) = \psi^\e \Big (x+
\frac{\e}{2} y\Big) \overline{\psi^\e} \Big(x- \frac{\e}{2} y\Big),
\end{align*}
as computed in \eqref{p-Fourier}. 
Inserting \eqref{middle} into $\widehat w^\e(x, y)$ we can write 
$$
|\langle \widehat w^\e  , \varphi \rangle  - \langle \rho^\e  , \varphi \rangle| \le |\langle R^\e, \varphi\rangle|,
$$
where for every $\varphi \in C_0(\R^{d}_x\times\R^d_y)$ the remainder $R^\e$ can be estimated using the Cauchy-Schwarz inequality:
\begin{align*}
\left | \iint_{\R^{2d}} R^\e(x,y) \varphi(x, y) \, dx \, dy \right | \le & \, \e^2 
\iint_{\R^{2d}} \varphi(x, y) \Big( \int_{-1}^1  \Big|\nabla \psi^\e \Big (x + \frac{\e s}{2}  y \Big ) \cdot y \Big| \, ds \Big)^2 dx \, dy \\
\le  & \, \e^2 C \iint_{\R^d}  \varphi(x,y) \Big | \nabla \psi^\e \Big (x + \frac{\e s}{2} \Big) y \Big|^2 dx  \, dy  \\
\le & \, \e^2 C \,  \| \nabla \psi^\e \|^2_{L^2(\Omega)},
\end{align*}
where the last inequality follows from a simple change of variables. 
We therefore conclude 
$\widehat w^\e(x,y)\rightharpoonup \rho(x)$, as $\e\rightarrow 0_+$ and an inverse Fourier transformation w.r.t. $y$ then yields the desired result. 
\end{proof}
\begin{remark} The proof given above, 
shows that the conclusion $\beta = \rho(x) \, \delta(p)$ still holds, if \eqref{consub} is replaced by the weaker assumption: $J^\e  \stackrel{\e\rightarrow 0_+}{\longrightarrow} 0 $, in $L_{\rm loc} ^1(\R^d)$ strongly.
\end{remark}

\subsection{The case of mono-kinetic Wigner measures} 

In situations where we have concentrations or oscillation on the critical scale $\e$ the connection between $\beta$ and $w$ is much more involved. The first problem we aim to analyze in more detail, is to find 
sufficient conditions under which the limiting Wigner measure is mono-kinetic. We remark that mono-kinetic Wigner measures correspond to the semi-classical limit of quantum dynamics before caustic onset time, see 
\cite{SMM} and Section 6 of this paper.

As a starting point in this direction we can state the following theorem, which 
can be seen as an analogue of Theorem \ref{thyoung1} for $\beta$. to this end, we recall the representation formula \eqref{reppsi}: For 
any $\psi^\e\in L^2(\R^d)$ we can write 
\begin{equation*}
\psi^\e(x) = \sqrt{\rho^\e(x)} e^{i S^\e(x)/ \e},
\end{equation*}
with $S^\e(x)\in \R$ defined $\rho^\e - a.e.$ (up to additive integer multiples of $2\pi$).

\begin{theorem} \label{theoremmono} Let $\psi^\e$ be uniformly bounded in $H^1_\e(\R^d)$, and assume $\rho^\e  \stackrel{\e\rightarrow 0_+
}{\longrightarrow}  \rho$ in $L^1(\R^d)$ strongly. If in addition there exists an extension of $S^\e$ to be denoted by the same symbol and a function $S\in C^1(\overline{\Omega})$, such that 
$$
\lim_{\e \to 0_+} \|\nabla S^\e  -  \nabla S\|_{L^\infty (\Omega)},
$$
where $\Omega  \subseteq \R^d$ is an open set containing ${\rm supp}\, \rho$, then it holds 
$$
w(x,p)=\beta(x,p)\equiv \rho(x) \, \delta (p- \nabla S(x)).
$$
\end{theorem}
\begin{proof} 
The assertion for $\beta$ follows immediately from the definition of $\beta^\e$, by using that $J^\e / \rho^\e = \nabla S^\e \stackrel{\e\rightarrow 0_+}{\longrightarrow} \nabla S$ 
uniformly on $\text{supp $\rho$}\subset \Omega$.

In order to prove the assertion for $w$, we use the $p$--Fourier transformed Wigner function and the representation \eqref{reppsi}, which yields
$$
\widehat w^\e(x, y ) = \sqrt{\rho^{\e} \Big (x + \frac{\e s}{2}  y \Big ) } \sqrt{\rho^{\e} \Big (x - \frac{\e s}{2}  y \Big )} \exp \left(i \delta S^\e(x,y) \right),
$$
where we denote the difference quotient
\begin{equation}\label{deltaS}
\delta S^\e(x,y) := \frac{1}{\e} \left(S^\e \Big (x + \frac{\e}{2}  y \Big )  -  S^\e \Big (x - \frac{\e}{2}  y \Big )\right).
\end{equation}
We aim to show that $\widehat w^\e$ converges weakly to 
$$
\widehat w(x, y ) = \rho(x) \exp \left(i y\cdot \nabla S(x)  \right).
$$
To this end we shall first show that $\sqrt{\rho^\e} \stackrel{\e\rightarrow 0_+}{\longrightarrow}  \sqrt{\rho}$ in $L^2(\R^d)$ strongly. Since, by assumption
$$
\int_{\R^d} \left( \sqrt{\rho^\e(x)} \right) ^2 dx \equiv  \int_{\R^d} {\rho^\e(x)} \, dx  \stackrel{\e\rightarrow 0_+
}{\longrightarrow}  \int_{\R^d} \rho(x)  \, dx=  \int_{\R^d} \left( \sqrt{\rho(x)} \right) ^2 dx ,
$$
it suffices to show $\sqrt{\rho^\e} \rightharpoonup \sqrt{\rho}$ in $L^2(\R^d)$ weakly. This, in turn, follows from a Young measure argument based on 
Proposition 4.2 and Proposition 4.3 of \cite{Pe1}. With the strong $L^2$ convergence at hand, we can write 
\begin{align*}
\widehat w^\e(x, y ) - \widehat w(x, y ) = & \, \left( \sqrt{\rho^{\e} \Big (x + \frac{\e }{2}  y \Big ) } \sqrt{\rho^{\e} \Big (x - \frac{\e}{2}  y \Big )} - \rho (x)\right) e^{i\delta S^\e(x) } \\
& \, + \rho(x)\left( e^{i \delta S^\e(x,y) }  - e^{i y\cdot \nabla S(x) }  \right).
\end{align*}
Due to the strong convergence of $\sqrt{\rho^\e}$ and the strong continuity of the shift-operator on $L^2(\R^d)$, 
the first term on the right hand becomes zero in the weak limit as $\e \to 0_+$, i.e. after localizing with a compactly supported test-function. It remains to estimate the 
third term, for which we use 
$$
\left | e^{i \delta S^\e(x,y)  }  - e^{i y\cdot \nabla S(x) }  \right| \le | \delta S^\e -  \nabla S \cdot y |  \le | \nabla S^\e -  \delta S| +|\delta S - \nabla S  |,
$$
by the mean-value theorem. Here, we first note that $|\delta S - \nabla S  | \stackrel{\e\rightarrow 0_+}{\longrightarrow} 0$, 
due to the assumed continuity of $\nabla S$. For the other term we again invoke the mean-value theorem and write 
$$
 | \nabla S^\e -  \delta S| \le \frac{\e}{2} \int_{-1}^1 y\cdot \nabla ( S^\e - S) \left (x+\frac{\e s}{2} y\right) \, ds.
$$
It is then easily seen that the this term likewise goes to zero, as $\e \to 0_+$, by assumption on (the gradient of) $S^\e$.
\end{proof}

\begin{remark} In view of Theorem \ref{thyoung1}, the above given assumptions are of course far from optimal when one is only concerned with the limit of $\beta$.\end{remark}

Alternatively, we can also show the following variant of Theorem \ref{theoremmono}, where 
we impose a slightly stronger assumption on the limiting phase $S$. In turn, the assumption on the extension $S^\e$ is slightly weaker than before.

\begin{corollary} Let $\Omega^\e  \subseteq \R^d$ be an open set containing $\text{supp $\rho^\e$}$. Then the assertion of Theorem \ref{theoremmono} also holds true, if 
there exists an extension $S^\e$ defined on $\Omega^\e$ and a function $S\in C^1\big(\overline{\bigcup _{\e\le1} \Omega^\e}\big)$ such that 
$$
\lim_{\e \to 0_+} \|\nabla S^\e  -  \nabla S\|_{L^\infty (\Omega^\e)}.
$$
\end{corollary}

\begin{proof}
The only difference from the proof given above is, that this time we write 
\begin{align*}
\widehat w^\e(x, y ) - \widehat w(x, y ) = & \, \sqrt{\rho^{\e} \Big (x + \frac{\e }{2}  y \Big ) } \sqrt{\rho^{\e} \Big (x - \frac{\e }{2}  y \Big )} e^{i y\cdot \nabla S(x) }  - \rho (x) e^{i y\cdot \nabla S(x) } \\
& \, + \sqrt{\rho^{\e} \Big (x + \frac{\e }{2}  y \Big ) } \sqrt{\rho^{\e} \Big (x - \frac{\e }{2}  y \Big )} \left( e^{i \delta S^\e(x,y)  }  - e^{i y\cdot \nabla S(x) }  \right).
\end{align*}
Due to the strong convergence of $\sqrt{\rho^\e}$ and the strong continuity of the shift-operator on $L^2(\R^d)$, 
the first two terms on the right hand side cancel each other in the limit $\e \to 0$ (see also Example III.1 in \cite{LiPa}). The second term can be treated similarly as before, using the mean-value theorem.
\end{proof}

\subsection{The general case} As we have seen, we cannot expect $w$ or $\beta$ to be mono-kinetic in general. It is therefore natural to 
study the connection between the two measures under more general circumstances. 
\begin{theorem}\label{thgeneral} 
Let $\psi^\e$ be uniformly bounded in $H^1_\e(\R^d)$ with corresponding densities $\rho^\e, J^\e\in L^1$. If 
$$ \e \nabla \sqrt{\rho^\e}\stackrel{\e\rightarrow 0_+
}{\longrightarrow} 0,\quad \text{\rm in $L^2_{\rm loc} (\R^d)$}
$$ 
and if there exists an extension of $S^\e$, such that
$$
\e \sup_{x \in \Omega^\e} \Big | \frac{\partial^2 S^\e } {\partial x_\ell \partial {x_j}} \Big |  \stackrel{\e\rightarrow 0_+
}{\longrightarrow} 0,\quad \forall \, \ell, j \in 1,\dots, d,
$$
where $\Omega^\e $ is an open set containing $\text{supp $\rho^\e$}$, then it holds 
$$
\lim_{\e \to 0_+} |\langle w^\e, \varphi \rangle - \langle  \beta^\e, \varphi \rangle| = 0, \quad \forall \, \varphi \in C_0(\R^d_x\times \R^d_p).
$$
\end{theorem} 
\begin{proof} We again consider $\psi^\e$ to be given via \eqref{reppsi}, and the corresponding $p$--Fourier transformed Wigner function
\begin{align*}
\widehat w^\e(x, y ) =   \sqrt{\rho^{\e} \Big (x + \frac{\e s}{2}  y \Big ) } \sqrt{\rho^{\e} \Big (x - \frac{\e s}{2}  y \Big )} \exp \left(i \delta S^\e(x,y) \right) ,
\end{align*}
which we want to compare with the $p$--Fourier transformed representation of $\beta^\e$, which, in view of \eqref{form}, is given by
$$
\widehat \beta^\e(x, y ) = \rho^\e(x) \exp \left(i y\cdot \nabla S^\e(x)  \right).
$$
To this end, we rewrite
\begin{align*}
\widehat w^\e(x, y ) =  \left(  \sqrt{\rho^{\e} \Big (x + \frac{\e s}{2}  y \Big ) } \sqrt{\rho^{\e} \Big (x - \frac{\e s}{2}  y \Big )} - \rho^\e(x) + \rho^\e(x) \right) \exp \left(i \delta S^\e(x,y) \right), 
\end{align*}
where $\delta S^\e$ is defined in \eqref{deltaS} and thus
\begin{align*}
 \delta S^\e(x,y) = & \,  \frac{1}{2} \int_{-1}^1 \nabla S^\e \Big (x\pm \frac{\e s}{2}  y \Big ) \cdot y \, ds \\
 = & \, \nabla S^\e(x) \cdot y +  \frac{\e^2}{2} \int_0^\tau \int_{-1}^1 y^\top D^2 S^\e \Big (x\pm \frac{\e s \tau }{2}  y \Big )  y \ d\tau \, ds ,
\end{align*}
with $D^2 S^\e$ denoting the Hessian matrix of $S^\e$. In other words, we have $$ \delta S^\e(x,y) =  \nabla S^\e(x) \cdot y  + \Phi^\e(x,y)$$ and thus we obtain
\begin{align*}
\widehat w^\e(x, y ) - \widehat \beta^\e(x, y )   = & \ \rho^\e(x)  e^{i \nabla S^\e(x) \cdot y} \left( e^{ i \Phi^\e(x,y) } - 1 \right)  \\
& \, + \left(  \sqrt{\rho^{\e} \Big (x + \frac{\e s}{2}  y \Big ) } \sqrt{\rho^{\e} \Big (x - \frac{\e s}{2}  y \Big )} - \rho^\e(x)  \right) e^{i \nabla S^\e(x) \cdot y}.
\end{align*}
In view of the assumption on $S^\e$ the first term on the right hand side goes to zero, as $\e \to 0_+$ and we therefore only need to take care of the second term. 
Using again the mean-value theorem we can rewrite 
\begin{align*}
&  \sqrt{\rho^{\e} \Big (x + \frac{\e s}{2}  y \Big ) } \sqrt{\rho^{\e} \Big (x - \frac{\e s}{2}  y \Big )} = \frac{\e}{2} \int_{0}^1 y\cdot 
 \nabla_z  \left( \sqrt{\rho^{\e}  (x + z ) }\right) \sqrt{\rho^{\e}  (x - z )}\Big |_{z = \e y/2} ds \\
& \,  +  \frac{\e}{2} \int_{0}^1 
\sqrt{\rho^{\e}  (x + z ) } \, y\cdot \nabla_z  \left(  \sqrt{\rho^{\e}  (x - z )} \right)  \Big |_{z = \e y/2} ds + \rho^\e(x).
\end{align*}
Now, let $\varphi \in C_0(\R^{2d})$ with $|\text{supp $\varphi$}| \le R < \infty$. Then, we can estimate, using the Cauchy-Schwarz inequality several times 
\begin{align*}
 & \,  \frac{\e}{2} \iint_{\R^{2d}} \int_{0}^1  \varphi (x,y) \sqrt{\rho^{\e}  (x \mp z )} \ y\cdot 
 \nabla_z  \left( \sqrt{\rho^{\e}  (x \pm z ) }\right) \Big |_{z = \e y/2} \, ds \, dx \, dy\\
 &\,  \le C(\varphi) \left( \int_{\R^d} \rho^\e(x) \, dx \right )^{1/2} \left( \int_{|x| < 2R} |\e \nabla \sqrt{\rho^\e(x)} |^2 \, dx \right )^{1/2} ,
 \end{align*}
 where $C(\varphi)>0$ depends on the $\text{supp $\varphi$}$. By assumption, this bound goes to zero as $\e \to 0_+$, which yields the assertion of the theorem.
\end{proof}

In Subsection \ref{sec:osc} we shall show that an $\e$-oscillatory velocity field $\nabla S^\e$ may cause the limiting Bohmian measure to be different from the Wigner measure. 
In view of \eqref{Ekin}, we also note that the condition $$ \e \nabla \sqrt{\rho^\e}  \stackrel{\e\rightarrow 0_+
}{\longrightarrow} 0 \quad \text{ in $L_{\rm loc} ^2(\R^d)$},$$ implies that the one part of the quantum mechanical kinetic energy which is not 
captured by the second moment of $\beta^\e$ has to converge to zero, at least locally in $x$. 
In fact, it is shown in the following corollary, that this is ``almost necessary" (i.e. at least for wave functions which are slightly more regular) to infer $\beta = w$.

\begin{corollary} Let $\psi^\e \in H^1_\e(\R^d)$ uniformly bounded as $\e \to 0_+$ and let $\e \nabla \psi^\e$ be compact at infinity. Furthermore assume that there exists a $\kappa>0$, such that
\begin{equation}\label{conmore}
|\e \xi|^{\kappa + 1} \widehat {\psi^\e(\xi)} \in L^2(\R^d),\quad \text{uniformly, as $\e \to 0_+$,}
\end{equation} 
and assume $w = \beta$. 
Then $ \e \nabla \sqrt{\rho^\e}  \stackrel{\e\rightarrow 0_+
}{\longrightarrow} 0$ in $L^2(\R^d)$. 
\end{corollary}

\begin{proof} Recall that 
\begin{equation}\label{energybal}
\begin{split}
\iint_{\R^{2d}} |p|^2 w^\e(x,p)  \, dx \, dp =  & \int _{\R^d} {|u^\e(x)|^2 \rho^\e(x)} \, dx +  \e^2 \int _{\R^d}  | \nabla \sqrt{\rho^\e(x)} |^2 \, dx \\
= & \iint_{\R^{2d}} |p|^2 \beta^\e (x,p) \, dx \, dp +  \e^2 \int _{\R^d}  | \nabla \sqrt{\rho^\e(x)} |^2 \, dx ,
\end{split}
\end{equation}
in view of \eqref{form} and \eqref{Ekin}. Thus
\begin{align*}
\iint_{\R^{2d}} |p|^2 \beta (x,p) \, dx \, dp \le & \,  \lim_{\e \to 0_+}\iint_{\R^{2d}} |p|^2 \beta^\e(x,p) \, dx \, dp \\
  \le & \, \lim_{\e \to 0_+} \iint_{\R^{2d}} |p|^2 w^\e(x,p) \, dx \, dp 
 =  \lim_{\e \to 0_+} \iint_{\R^{2d}} |p|^2 n^\e (p) \, dp ,
\end{align*}
where $n^\e$ denotes the momentum density, i.e.
$$
n^\e(p):=  \int_{\R^{d}} w^\e(x,p) \, dx = \e^{- d} \left |\widehat {\psi^\e}\left(\frac{p}{\e} \right) \right |^2.
$$
Now, using the results given in \cite[Proposition 1.7]{LiPa}, it is easy to show that
$$
 \lim_{\e \to 0_+} \iint_{\R^{2d}} |p|^2 w^\e(x,p) \, dx \, dp=   \iint_{\R^{2d}} |p|^2 w(x,p) \, dx \, dp, 
$$
provided that $\psi^\e$ satisfies the assumptions stated above. Since 
$\beta = w$, by assumption, we obtain
$$
 \lim_{\e \to 0_+} \iint_{\R^{2d}} |p|^2 \beta^\e(x,p) \, dx \, dp =  \iint_{\R^{2d}} |p|^2 w(x,p) \, dx \, dp = \iint_{\R^{2d}} |p|^2 \beta(x,p) \, dx \, dp .
$$ 
We therefore conclude from \eqref{energybal} that $ \e \nabla \sqrt{\rho^\e}  \stackrel{\e\rightarrow 0_+
}{\longrightarrow} 0$ in $L ^2(\R^d)$.
\end{proof}

In combination with Theorem \ref{thgeneral} we conclude that for wave functions $\psi^\e$ which are uniformly bounded in any $\e$-scaled Sobolev space of higher order than $H^1_\e(\R^d)$ and for 
which $\e \nabla \psi^\e$ is compact at infinity, 
the fact that $ \e \nabla \sqrt{\rho^\e}  \stackrel{\e\rightarrow 0_+
}{\longrightarrow} 0$ in $L_{\rm loc} ^2(\R^d)$, is indeed a necessity to obtain $w = \beta$. 

\begin{remark} Note that condition \eqref{conmore} is trivially 
propagated by the free Schrödinger dynamics corresponding to $V(x) =0$. Moreover,  if $V(x)$ satisfies $\partial^\alpha V \in L^\infty(\R^d)$ for all $ |\alpha| \le 2$, 
a simple Grownwall estimate, combined with energy and mass conservation, shows that \eqref{conmore} with $\kappa \in [0,1]$ is propagated by the 
Schrödinger dynamics $U^\e(t) = e^{-it H^\e /\e}$ on bounded time-intervals.
\end{remark}

\section{Case studies} \label{cases} In this section we shall study the case of oscillations and concentration effects on the (critical) scale $\e$ and compare the corresponding Bohmian and Wigner measure. 
In general we can expect all of these effects to be physically relevant, see e.g. the examples given in \cite{GaMa, SMM}. 

\subsection{Oscillatory functions}\label{sec:osc} Let $\psi^\e(x) = f(x) g\left(\frac{x}{\e}\right)$, where $f \in C_0^\infty(\R^d; \C)$ and $g \in C^\infty(\R^d, \C)$ is assumed to be \emph{periodic} w.r.t. some lattice $L\simeq \mathbb Z^d$, i.e. 
$g(y+ \ell) = g(y)$ for any  $y\in \R^d$ and $\ell \in L$. In other words, $\psi^\e$ is a slowly modulated high-frequency oscillation. Computing the corresponding Bohmian measure we find
$$
\beta^\e(x,p) = |f(x)|^2 \left| g\left(\frac{x}{\e}\right)\right|^2 \, \delta \left(p - \im \left(\frac{\nabla g \left(\frac{x}{\e}\right)}{g \left(\frac{x}{\e}\right)}+ \e \frac{\nabla f (x)}{f (x)}\right) \right).
$$
Taking $\varphi \in C_0(\R^d_x \times \R^d_p)$ we conclude by invoking the theory of \emph{two-scale convergence} (see e.g. \cite{Al}), that 
\begin{equation*}
\langle  \beta^\e, \varphi \rangle  \stackrel{\e\rightarrow 0_+ }{\longrightarrow} \frac{1}{|Y|}  \int_{\R^d} \int_{Y}  |f(x)|^2  |g(y)|^2 \varphi \left(x, \im \left( \frac{\nabla g(y)}{g(y)} \right)\right) \, dy \, dx,
\end{equation*}
where $Y \subset L$ is the fundamental domain of the lattice $L$. We thus find, that the limiting Bohmian measure is given by
\begin{equation}\label{oslimit}
\beta(x,p) = |f(x)|^2 \, \frac{1}{|Y|} \int_{Y} |g(y)|^2 \ \delta\left(p- \im \left( \frac{\nabla g(y)}{g(y)} \right ) \right) dy.
\end{equation}
On the other hand, let 
$$
g(y) = \sum_{\ell^* \in L^*} \hat g_{\ell ^*} e^{-i y\cdot  \ell^*},
$$
be the Fourier series of $g(y)$, where $L^*$ denotes 
the corresponding dual lattice, and consider the Wigner function of $\psi^\e$, after Fourier transformation w.r.t. the variable $p\in \R^d$, i.e.
$$
\widehat w^\e(x, y) = f \Big (x+
\frac{\e}{2} y\Big) \overline{f} \Big(x- \frac{\e}{2} y\Big) \sum_{\ell^*, m^* } \hat g_{\ell ^*} \overline {\hat g}_{m ^*}e^{-i ((x/ \e +  y/2) \cdot  \ell^* - (x/ \e -  y/2) \cdot  m^*))}.
$$
Then it is easy to see that, as $\e \to 0_+$:
$$
\widehat w^\e(x, y) \rightharpoonup |f(x)|^2 \sum_{\ell^* \in L^*} |\hat g_{\ell ^*} |^2 e^{i y\cdot  \ell^*}.
$$
More precisely we find that
\begin{equation*}
\langle w^\e, \varphi \rangle  \stackrel{\e\rightarrow 0_+ }{\longrightarrow} \sum_{\ell^* \in L^*} |\hat g_{\ell ^*} |^2 \int_{\R^d} |f(x)|^2 \varphi (x, \ell^*) dx,
\end{equation*}
and hence the Wigner measure associated to the $L$-oscillatory function $\psi^\e$ is given by 
\begin{equation}\label{oslimitw}
w(x,p) = |f(x)|^2 \,   \sum_{\ell^* \in L^*} |\hat g_{\ell ^*} |^2 \delta (p - \ell^*),
\end{equation}
which should be compared to \eqref{oslimit}. 
\begin{lemma}\label{lemosc}
For $g \in C^\infty (\R^d)$ the limiting measures \eqref{oslimit} and \eqref{oslimitw} coincide, if and only if  $g(x)$ carries only a single oscillation $\ell^* \in L^*$.
\end{lemma}

\begin{proof} In order to show that $\beta \not = w$ it is enough to prove that their respective second moments do not coincide. To this end, we compute
$$
\int_{\R^{d}} |p|^2 \, \beta^\e(x, dp) = |Y|^{-1} |f(x)|^2 \int_{Y} |g(y)|^2 \left| \im \left( \frac{\nabla g(y)}{g(y)} \right) \right|^2 dy
$$
and
$$
\int_{\R^{d}} |p|^2 \, w^\e(x, dp) = |f(x)|^2 \sum_{\ell^* \in L^*} |\ell^*|^2  |\hat g_{\ell^*}|^2 \equiv |Y|^{-1} |f(x)|^2 \int_{Y}  |\nabla g(y)|^2 dy.
$$
Using the polar decomposition $g(y) = r(y) e^{ i \theta(y)}$ these integrals can be rewritten as
$$
\int_{Y} |g(y)|^2 \left| \im \left( \frac{\nabla g(y)}{g(y)} \right) \right|^2 dy = \int_{Y} |r(y)|^2 |\nabla \theta (y)|^2 dy 
$$
and
$$
 \int_{Y}  |\nabla g(y)|^2 dy = \int_{Y} |\nabla r(y)|^2 + |r(y)|^2 |\nabla \theta (y)|^2 dy.
$$
Obviously these two integrals can only coincide, if $|\nabla r(y)| = 0$, which implies $g(y) = c e^{i \theta(y)}$, with $c \ge0$ and $\theta(y) \in \R$. 
In this case the support of the $x$-projection of $\beta$ is the closure of the range of $\nabla \theta$, i.e. bounded in $\R^d_p$. On the other hand, the support of the $x$-projection of $w$ is $L^*$.
Hence, for a smooth function $g$ the two supports can only be equal if $\theta(y) = y\cdot \ell^*$ for some $\ell^*\in L^*$, in which case $w=\beta$.
\end{proof}

Assume now that $f$ is real-valued and let $g(y) = e^{i \theta (y)}$. Then, the sequence $\psi^\e$ is obviously uniformly bounded in $H^1_\e(\R^d)$ and the phase $S^\e(x)= \e \theta(x/ \e)$ is such that 
$$ \e \, \frac{\partial^2 S^\e } {\partial x_\ell \partial {x_j}} =  \frac{\partial^2 \theta } {\partial y_\ell \partial {y_j}} \left( \frac{x}{\e} \right ),\quad \ell, j =1,\dots , d.$$ 
Therefore the first assumption of Theorem \ref{thgeneral} is satisfied, but the second is not, unless $\theta = 0$. As stated above, $\beta \not = w$.

\subsection{Concentrating functions} We consider wave function $\psi^\e$ which concentrate at a single point. To this end, let, for some $x_0\in \R^d$, $\psi^\e(x) = \e^{-{d/2}} f \left(\frac{x-x_0}{\e}\right)$ with 
$f \in C_0^\infty(\R^d; \C)$. Thus $|\psi^\e(x)|^2  \rightharpoonup \delta(x-x_0)$, as $\e \to 0_+$. The corresponding Wigner measure has been already computed in \cite{Ge, LiPa} as
\begin{equation}\label{colimitw}
w(x,p)= (2\pi)^{-d}  |\widehat f(p)|^2  \, \delta(x-x_0),
\end{equation}
where $\widehat f$ denotes the Fourier transform of $f$. On the other hand, we easily compute
$$
\beta^\e(x,p) = \e^{-d} \left | f \left(\frac{x}{\e}\right)\right |^2 \, \delta \left(p - \im \left(\frac{\nabla f \left (\frac{x-x_0}{\e}\right)}{f \left(\frac{x-x_0}{\e}\right)} \right) \right),
$$
and thus
\begin{align*}
\langle  \beta^\e, \varphi \rangle & = \ \e^{-d} \int_{\R^d}   \left | f \left(\frac{x}{\e}\right)\right |^2 \varphi \left(x, \im \left(\frac{\nabla f \left (\frac{x-x_0}{\e}\right)}{f \left(\frac{x-x_0}{\e}\right)} \right) \right )dx \\
& = \  \int_{\R^d}   | f (p) |^2 \varphi \left(\e p + x_0, \im \left(\frac{\nabla f  (y)} {f (p)} \right)\right) dp ,
\end{align*}
by a simple change of variables. We therefore conclude
\begin{equation*}
\langle  \beta^\e, \varphi \rangle  \stackrel{\e\rightarrow 0_+ }{\longrightarrow}    \int_{\R^d}   | f (p) |^2 \varphi \left( x_0, \im \left(\frac{\nabla f  (p)} {f (p)} \right)\right) dp .
\end{equation*}
In other words,
\begin{equation}\label{colimit}
\beta(x,p) = \delta(x-x_0) \,   \int_{\R^d}   | f (y) |^2 \delta\left(p-\im \left(\frac{\nabla f  (y)} {f (y)} \right)\right) dy, 
\end{equation}
Again we see that the Wigner measure \eqref{colimitw} and the classical limit of the Bohmian measure \eqref{colimit} are rather different in this case. 
\begin{lemma}
The limiting measures \eqref{colimit} and \eqref{colimitw} do not coincide, unless $f=0$.
\end{lemma}

\begin{proof}
Again, it is enough to prove that the second moments of $\beta$ and $w$ do not coincide. By the same arguments as in the proof of Lemma \ref{lemosc}, we conclude that the second moments 
can only coincide if $f(y) = c e^{i \theta(y)}$, $c\ge 0$, which is in contradiction to the fact that $f \in L^2(\R^d)$, unless $c=0$.
\end{proof}
\subsection{Examples from quantum physics}  As a possible application we shall now consider some particular examples of quantum mechanical wave functions, which incorporate 
oscillatory and concentrating effects in their classical limit.

\begin{example}[Semi-classical wave packets] In this example we consider so-called semi-classical wave packets (or coherent states), 
which incorporate, both, oscillations and concentrations, i.e. 
$$
\psi^\e(x) = \e^{-d/4}  f \left(\frac{x-x_0}{\sqrt{\e}}\right) e^{i p_0 \cdot x/ \e}, \quad x_0, p_0 \in \R^d,
$$
for some given profile $f \in C_0^\infty(\R^d; \C)$. Similarly as before, we compute
\begin{align*}
\langle  \beta^\e, \varphi \rangle  = \ \e^{-d/2} \int_{\R^d}   \left | f \left(\frac{x-x_0}{\sqrt{\e}}\right)\right |^2 \varphi \left(x, p_0+\sqrt{\e} \, \im \left(\frac{\nabla f \left (\frac{x-x_0}{\sqrt{\e}}\right)}{f \left(\frac{x-x_0}{\sqrt{\e}}\right)} \right) \right )dx ,
\end{align*}
which in the limit $\e\to0_+$ yields
$$
\beta(x,p) =  \int_{\R^d} | f(x)|^2 dx\, \delta(x-x_0) \, \delta(p-p_0).
$$
On the other hand, the Wigner measure of a coherent state is found in \cite{LiPa} to be
$$
w(x,p)= \int_{\R^d} | f (x)|^2 dx\, \delta(x-x_0)\, \delta(p-p_0).
$$
Thus, $\beta = w$ in this case, a fact which makes coherent states particularly attractive for the study of the classical limit of Bohmian dynamics \cite{DuRo}. 
Note that for $p_0 =0$ this can be seen as a particular case of Theorem \ref{thsub}, since coherent states concentrate on the scale $\sqrt{\e}$.
\end{example}

\begin{example}[Eigenfunctions] 

Let us consider a Hamiltonian operator
$$H^\e=-\frac{\varepsilon^2}{2}\Delta+V(x),$$ 
with (real-valued) smooth confining potential $V(x)\to+\infty$ as $|x|\to\infty$. The corresponding spectrum is known to be discrete and the associated spectral problem reads
$$
H^\e \psi_n^\e = \lambda_n^\e \ \psi_n^\e, \quad n \in \N,
$$
with normalized eigenstates $\psi^\e_n\in L^2(\R^d)$ and eigenvalues
$\lambda_n\in \R$. Now, let $\{\e_n \}_{n \in \N}$ be a sequence such that $\e_n \stackrel{n \rightarrow \infty }{\longrightarrow}0$ and $\lambda_n^{\e_n}  \stackrel{n \rightarrow \infty }{\longrightarrow} \Lambda\in \R$. 
Since $V(x)$ is confining (and since $\psi_n^{\e_n}$ is normalized) there exists a subsequence, which we denote by the same symbol, such that 
$$
\big|\psi_n^{\e_n}\big|^2  \stackrel{n \rightarrow \infty }{\longrightarrow} \rho(x)\not = 0,
$$
weakly in measure. Since $H^\e$ is self-adjoint the eigenfunctions $\psi_n^\e$ can be chosen real-valued and we therefore conclude $\beta(x,p) = \rho(x)\, \delta(p)$.

For the particular case of the harmonic oscillator $V(x) = \frac{1}{2}|x|^2$ the Wigner measure $w$, as computed in \cite{LiPa}, is 
$w(x,p) = \delta(|x|^2 + |p|^2 - \Lambda)$, i.e. 
a uniform distribution on the energy sphere. Thus, $w \not = \beta$, unless $\Lambda = 0$ and $\rho(x)=\delta(x)$, which corresponds to a classical particle at rest, sitting at the minimum of $V(x)=\frac{1}{2}|x|^2$. 
In more generality, the fact that $w\not = \beta$ can be concluded by invoking results from quantum ergodicity, see e.g. \cite{HMR} or microlocal analysis, which shows that 
$${\rm supp} \, w\subseteq \big \{x,p \in \R^{2d}: \, \textstyle{\frac{1}{2}} |p|^2+V(x)=\Lambda \big \}.$$
\end{example}

\section{Connection to WKB approximations} \label{sec:WKB}

WKB expansions are a standard approach in semi-classical approximation of quantum mechanics (see e.g. \cite{Ca1, SMM} and the references given therein). To this end one seeks an approximation of the 
exact solution $\psi^\e(t,x)$ to \eqref{schro}, in the following form
\begin{equation}\label{ansatz}
\psi_{\rm wkb}^\e(x)= a^\e(t,x) e^{i S(t, x) /\e},
\end{equation}
where $S(t,x)\in \R$ is some $\e$-independent (real-valued) phase function and $a^\e(x)$ a slowly varying amplitude (not necessarily real-valued), which admits an asymptotic expansion 
$$a^\e \sim a + \e a_1+ \e^2 a_2+\dots.$$
Note that the ansatz \eqref{ansatz} specifies a certain 
$\e$-oscillatory structure of $\psi^\e$ due to the fact that the phase $S(x)$ is assumed to be $\e-$\emph{independent}. In particular, 
it should \emph{not} be confused with the representation \eqref{reppsi}. 
Obviously, we find that the Bohmian measure in this case is given by
$$
\beta^\e[\psi_{\rm wkb}^\e(t)] = |a^\e(t,x)|^2\, \delta (p- \nabla S(t,x)).
$$
Plugging 
\eqref{ansatz} into the Schr\"odinger equation \eqref{schro} and assuming sufficient smoothness, one obtains in leading order the following equation for the 
the phase
\begin{equation}\label{hj}
\partial_t S + \frac{1}{2} |\nabla S|^2 + V(x)= 0
\end{equation}
and the leading order amplitude 
\begin{equation}\label{wkbamp}
\partial_t a + \nabla a \cdot \nabla S + \frac{a}{2} \Delta S= 0,
\end{equation}
which is easily rewritten as a conservation law for $\rho=a^2$, i.e. 
$$
\partial_t \rho + \diver (\rho \nabla S) = 0.
$$
Equation \eqref{hj} is nothing but the classical \emph{Hamilton-Jacobi equation}. If we set $u= \nabla S$, then we clearly obtain from \eqref{hj}
the inviscid field-driven Burgers equation
\begin{equation}\label{burger}
\partial_t u + (u\cdot \nabla) u + \nabla V (x) = 0,
\end{equation}
which can formally be seen as the classical limit of \eqref{ueq}. 

The main problem of the WKB approach arises from the fact that \eqref{hj}, or equivalently \eqref{burger}, in general does not admit global smooth solutions. 
In general $S(t,\cdot)\in C^\infty(\R^d)$ only for $t \in [0, T^*)$, for some (typically small) finite time $T^*> 0$, which marks the appearance of the the first \emph{caustic}, or, equivalently, the appearance of 
the first shock in \eqref{burger}, cf. \cite{Ca1, SMM}. Caustics reflect the fact that new $\e$-scales are generated in the exact solution $\psi^\e(t,x)$, which are no longer captured by the 
simple ansatz \eqref{ansatz}. Nevertheless, at least locally in-time the WKB approximation yields a simple representation for $\psi^\e(t,x)$ which can be extended to the case of 
nonlinear Schr\"odinger equations, see \cite{Ca1, Ca2}. Its connection to Wigner measures has been extensively studied in \cite{SMM}. 
The connection to Bohmian measures is given in the following result.

\begin{proposition} Let Assumption \ref{assumptionV} and \ref{assumptionW} hold and let $T^*>0$ be the caustic onset time. Assume there exist smooth solutions $a, S \in C^\infty([0,T^*)
\times \R^d)$, with $a(t,\cdot) \in L^2(\R^d)$. Then, for the exact solution $\psi^\e(t)$ of the Schr\"odinger equation with WKB initial data, it holds 
$$
\beta(t,x,p)=w(t,x,p)\equiv |a(t,x)|^2 \, \delta(p-\nabla S(t,x)), \quad  \forall \, t \in [0, T^*).
$$
\end{proposition}
\begin{proof} The statement for the Wigner measure has been proven in \cite{GaMa} and (in more generality also in \cite{SMM}). 
In order to prove that $\beta$ is mono-kinetic before caustics we refer to \cite{AlCa}, where it is shown that for $ T \in [0, T^*)$:
$$
\rho^\e \stackrel{\e\rightarrow 0_+
}{\longrightarrow} |a|^2,\quad 
J^\e \stackrel{\e\rightarrow 0_+
}{\longrightarrow} |a|^2 \nabla S,
$$
in $C([0,T]; L^1(\R^d))$ strongly. 
Thus, recalling Theorem \ref{thyoung1}, we directly conclude the desired result. 
\end{proof}

In other words, as long as the WKB approximation is valid (i.e. locally in-time before caustics) the classical limit of the Bohmian measure of the \emph{true solution} to the Schr\"odinger equation is mono-kinetic 
and the same holds for the Wigner measure. For the latter it has been shown in \cite{SMM} that locally away from caustics the Wigner measure can always be written as a sum of mono-kinetic terms. 
The proof requires the use of the Hamiltonian flow \eqref{hamode}
governing $w(t)$. Unfortunately, such a limiting phase space flow is not available for $\beta(t)$. All we can conclude from above is that for $t \in [0, T^*)$, the dynamics of $\beta(t)$ is governed by 
\begin{equation}
\left \{
\begin{split}
& \,  \dot X= P , \quad X(0,x) = x, \\
& \, \dot P= - \nabla V(X) , \quad P(0,x) = \nabla S(0, x)\equiv u(x).
\end{split}
\right. 
\end{equation}
This is the characteristic flow associated to \eqref{hj}. Since it breaks down at caustics no information for $t\ge T^*$ can be obtained by following this approach. In view of the examples given in Section \ref{cases} 
and the already known concentration and oscillation effects beyond caustics (see e.g. \cite{GaMa, SMM}) we cannot expect a simple description of the classical limit of Bohmian trajectories in this case.
A possible way to overcome this 
problem could be a Young measure analysis of the Bohmian trajectories in the spirit of \cite{AKST} (see also Remark \ref{rem: measXP}), which, however, is beyond the scope of this work. 
We also note that in the case of of semi-classical wave packets treated in \cite{DuRo}, the problem of caustics does not appear.

\begin{remark}
In order to give the reader a basic intuition on the limiting behavior of the Bohmian measure after caustics, we recall that by stationary phase arguments (see e.g. \cite{SMM}) the 
wave function after caustics can be approximated by a superposition of WKB states. To illustrate the kind of
phenomena which can happen in this situation,
 we consider here a sum of two WKB states, i.e. 
\begin{equation*}
\psi^\e(x) =a_1 e^{i S_1(x) / \e}+ a_2 e^{i S_2(x) / \e},
\end{equation*}
with real-valued $a_1, a_2 \in C^\infty_0(\R^d)$, $S_1, S_2 \in C^\infty(\R^d)$, such that, for all $x\in \R^d$ it holds: $\nabla S_1(x)\not = \nabla S_2(x)$ and $a_1 (x) > a_2(x)$. 

One the one hand, we infer from \cite{SMM}, that, in this case the Wigner measure  is given by 
$$
w(x,p)= a_1^2(x) \, \delta (p - \nabla S_1(x)) + a_2^2(x) \, \delta (p - \nabla S_2(x)),
$$
i.e. the sum of two mono-kinetic measures. On the other hand, a lengthy but straightforward computation shows that the limiting Bohmian measure is given by
\begin{equation}\label{bik}
\beta(x,p)=\frac{1}{2\pi}  \int_{0}^{2\pi} n(x,\theta)\, \delta(p- \Phi (x,\theta))\, d\theta,
\end{equation}
where
\begin{equation*}
n(x,\theta):= a^2_1(x)+ a^2_2(x) + 2 a_1(x) a_2(x) \cos{\theta},
\end{equation*}
and 
\begin{equation*}
\Phi:=\frac{1}{n(x,\theta)} \left( 
a_1^2(x) \nabla S_1(x)+ a^2_2(x)\nabla S_2(x)+ a_1 (x)a_2 (x) \cos{\theta} (\nabla S_1(x)+
\nabla S_2(x)\Big)\right).
\end{equation*}
To this end, we note that the computation of the current vector field $ \im \left(\frac{ \nabla \psi^\e(x)}{ \psi^\e(x)} \right )$ yields a smooth function which is periodic in $\theta(x) = (S_2(x)-S_1 (x))/\e$ and thus admits a 
Fourier expansion w.r.t. $\theta$. By standard two scale-convergence we infer that the limit as $\e \to 0_+$ is given by the zeroth order coefficient of this Fourier series, from which 
we deduce (\ref{bik}). 

Finally, let us mention that multi-phase type WKB methods have been used recently, 
for the study of the ``quantum hydrodynamic" regularisation of the Burgers equation \cite{P} (see also \cite{GaMa, SMM}).

\end{remark}

\textbf{Acknowldegment.} The authors want to thank Wilfrid Gangbo for stimulating discussions on the derivation of \eqref{kinetic} and on optimal transportation formulations of the Schr\"odinger equation.


\begin{thebibliography}{amsplain}

\bibitem{AlCa} T. Alazard and R. Carles, \emph{Supercritical geometric optics for nonlinear Schrödinger equations}.
Arch. Ration. Mech. Anal. {\bf 194} (2009), no. 1, 315--347.

\bibitem{Al} G. Allaire, \emph{Homogenization and two-scale convergence}. SIAM J. Math. Anal. {\bf 23} (1992), 1482--1518.  

\bibitem{Am} L. Ambrosio, \emph{Transport Equation and Cauchy Problem for Non-Smooth Vector Fields}. In: Calculus of Variations and Nonlinear Partial Differential Equations, 
Springer Lecture Notes in Mathematics 1927, Springer 2008.

\bibitem{AMGA} L. Ambrosio and W. Gangbo, \emph{Hamiltonian ODE's in the Wasserstein space of probability measures}. Comm. Pure Applied Math. {\bf 61} (2008), 18--53.

\bibitem{AKST} Z. Artstein, I. G. Kevrekidis, M. Slemrod, and E. S. Titi, \emph{Slow observables of singularly perturbed differential equations}, Nonlinearity {\bf 20} (2007), 2463--2481.

\bibitem{Ba} J. M. Ball, \emph{A version of the fundamental theorem for Young measures}. In: PDEs 
and Continuum Models of Phase Transitions, Lecture Notes in Physics, Vol. 
344, Rascle, M., Serre, D., Slemrod, M. (eds.), Springer 1989.

\bibitem{BDGPZ} K. Berndl, D. Dürr, S. Goldstein, G. Peruzzi, and N. Zanghi, \emph{On the global existence of Bohmian mechanics}. Comm. Math. Phys. {\bf 173} (1995), 647--673.

\bibitem{BeMe}  J. A. Beswick and C. Meier, \emph{Hybrid Quantum/Classical Dynamics using Bohmian trajectories}. Chemical Physics Series vol. 83, Springer Verlag (2006). 


\bibitem{BO} D. Bohm, \emph{A Suggested Interpretation of the Quantum Theory in Terms of ``Hidden Variables" I, II}. Phys. Rev. {\bf 85} (1952), 166--193.

\bibitem{Ca1} R. Carles, \emph{Semi-classical analysis for nonlinear Schr\"odinger equations}.
World Scientific, Co. Pte. Ltd., Hackensack, NJ 2008.

\bibitem{Ca2} R. Carles, \emph{WKB analysis for nonlinear Schr\"odinger equations with potential}.
Comm. Math. Phys. 269 (2007), no. 1, 195-221.

\bibitem{DDP} D. A. Deckert, D. Dürr, and P. Pickl, \emph{Quantum Dynamics with Bohmian Trajectories}. J. Phys. Chem. A {\bf 111} (2007), 10325--10330.

\bibitem{DiLi} R. J. DiPerna and P. L.  Lions, \emph{Ordinary differential equations, transport theory 
and Sobolev spaces}. Invent. Math. {\bf 98} (1989), 511--547.

\bibitem{DuTe} D. D\"urr and S. Teufel, \emph{Bohmian Mechanics}. Springer Verlag, 2009.

\bibitem{DuRo} D. Dürr S. Römer, \emph{On the classical limit of Bohmian mechanics for Hagedorn wave packets}. Preprint {\tt  arXiv:1003.5159}.

\bibitem{GaMa} I. Gasser and P. A. Markowich, \emph{Quantum hydrodynamics, Wigner transforms and the classical limit}. Asympt. Anal. {\bf 14} (1997), 97--116.

\bibitem{GNT} W. Gangbo, T. Nguyen, and A. Tudorascu, \emph{Hamilton-Jacobi equations in the Wasserstein space}. Methods Appl. Anal. {\bf 15} (2008), no. 2, 155--183. 

\bibitem{Ge} P. G\'erard, \emph{Mesures semi-classiques et ondes de Bloch}, Expos\'e de l'Ecole Polytechnique, E.D.P., Expos\'e no. XVI, (1991). 

\bibitem{GMMP}  P. G\'erard, P. A. Markowich, N. J. Mauser, and F. Poupaud, \emph{Homogenisation Limits and Wigner transforms}. Comm. Pure Appl. Math.
\textbf{50} (1997), 323--379.

\bibitem{GST} U. Gianazza, G. Savar\'e, and G. Toscani, \emph{The Wasserstein gradient flow of the Fisher information and the Quantum Drift-Diffusion equation.} 
Archive Rational Mech. Anal. {\bf 194} (2009), no. 1, 133--220.

\bibitem{GMB} E. Gindensperger, C. Meier, and J. A. Beswick, \emph{Mixing quantum and classical dynamics using Bohmian trajectories.} J. Chem. Phys. {\bf 113} (2000), issue 21, 9369--9372.

\bibitem{HMR} B. Helffer, A. Martinez, and D. Robert, \emph{Ergodicit\'e et limite semi-classique}. Comm. Math. Phys. {\bf 109} (1987), 313--326.

\bibitem{HoEa} R. W. Hockney and J. W. Eastwood, \emph{Computer simulation using particles}, Institute of Physics (1988).

\bibitem{LiPa}   P. L. Lions and T. Paul, \emph{Sur les measures de
Wigner}. Rev. Math. Iberoamericana \textbf{9} (1993), 553--618.

\bibitem{MA} E. Madelung, \emph{Quanten Theorie in hydrodynamischer Form}, Zeitschrift f\"ur Physik, {\bf 40} (1926), 322--326.

\bibitem{NeFre} D. Nerukh and J. H. Frederick, \emph{Multidimensional quantum dynamics with trajectories: a novel numerical implementation of Bohmian mechanics}. 
Chem. Phys. Lett. {\bf 332} (2000), issue 1-2, 145--153.

\bibitem{P} T. Paul, \emph{Some remarks concerning the  Burgers equation and quantum hydrodynamics}, preprint.

\bibitem{Pe1} P. Pedregal, \emph{Optimization, relaxation and Young measures}, Bull. Amer. Math. Soc. {\bf 36} (1999), no. 1, 27--58.

\bibitem{Pe2} P. Pedregal, \emph{Parametrized Measures and Variational Principles}, Birkh\"auser, Basel 1997.

\bibitem{ReSi} M. Reed and B. Simon, \emph{Methods of Modern Mathematical Physics II}, Academic Press (1975).

\bibitem{SMM} C. Sparber, P.~Markowich, and N.~Mauser, {\it Wigner functions vs. WKB methods in multivalued geometrical optics}. Asymptot. Anal. {\bf 33} (2003), no. 2, 153--187.

\bibitem{TeTu} S. Teufel and R. Tumulka, \emph{Simple proof of global existence of Bohmian trajectories}. Comm. Math. Phys. {\bf 258} (2005), 349--365.

\bibitem {Wi}  E. Wigner, \emph{On the quantum correction for the
thermodynamical equilibrium}. Phys. Rev. \textbf{40} (1932), 742--759.

\bibitem{WKH} R. E. Wyatt, D. J. Kouri, and D. K. Hoffman, \emph{Quantum wave packet dynamics with trajectories: Implementation with distributed approximating functionals}. J. Chem. Phys. {\bf 112} (2000), 10730--10738.

\bibitem{WyTr} R. E. Wyatt and C. J. Trahan, \emph{Quantum dynamics with trajectories: introduction to quantum hydrodynamics}. Springer, Berlin (2005).  

\end{thebibliography}
\end{document}